\documentclass[twocolumn]{aastex63}
\usepackage{mathptmx}
\usepackage{graphicx} 
\usepackage{threeparttable} 
\usepackage[caption=false]{subfig}
\usepackage{graphicx}
\usepackage{multirow}
\usepackage{graphicx}
\usepackage{booktabs}
\usepackage{amsmath, amsthm, amssymb, amsfonts}
\usepackage[T1]{fontenc}
\usepackage{epstopdf}
\graphicspath{{./}{figures/}}

\usepackage{bm}
\begin{document}

\title{Origin of the Very High Energy $\gamma$-rays in the Low-luminosity Active Galactic Nucleus NGC 4278}

\correspondingauthor{Jin Zhang}
\email{j.zhang@bit.edu.cn}

\author[0000-0003-2547-1469]{Ji-Shun Lian}
\affiliation{School of Physics, Beijing Institute of Technology, Beijing 100081, People's Republic of China; j.zhang@bit.edu.cn}

\author[0009-0003-9471-4724]{Jia-Xuan Li}
\affiliation{School of Physics, Beijing Institute of Technology, Beijing 100081, People's Republic of China; j.zhang@bit.edu.cn}

\author{Xin-Ke Hu}
\affiliation{School of Physics, Beijing Institute of Technology, Beijing 100081, People's Republic of China; j.zhang@bit.edu.cn}

\author[0000-0002-4789-7703]{Ying-Ying Gan}
\affiliation{School of Physics, Beijing Institute of Technology, Beijing 100081, People's Republic of China; j.zhang@bit.edu.cn}

\author[0009-0006-4551-8235]{Tan-Zheng Wu}
\affiliation{School of Physics, Beijing Institute of Technology, Beijing 100081, People's Republic of China; j.zhang@bit.edu.cn}

\author[0000-0001-6863-5369]{Hai-Ming Zhang}
\affiliation{Guangxi Key Laboratory for Relativistic Astrophysics, School of Physical Science and Technology, Guangxi University, Nanning 530004, People's Republic of China}

\author[0000-0003-3554-2996]{Jin Zhang\dag}
\affiliation{School of Physics, Beijing Institute of Technology, Beijing 100081, People's Republic of China; j.zhang@bit.edu.cn}

\begin{abstract}

NGC 4278, a Low-luminosity active galactic nucleus (AGN), is generally classified as a low-ionization nuclear emission line region (LINER). Recently, it has been reported to be associated with a very high energy $\gamma$-ray source 1LHAASO J1219+2915 in the first Large High Altitude Air Shower Observatory source catalog. However, no associated counterpart has been detected by analyzing the data collected by the Large Area Telescope on board the Fermi Gamma-ray Space Telescope. By analyzing its X-ray observation data from {\it Swift}-XRT, we find NGC 4278 is in a high-flux state on MJD 59546, with the X-ray flux more than one order of magnitude higher than that observed $\sim$ 11.7 year earlier by {\it Chandra}. Interestingly, this {\it Swift}-XRT observation was conducted during the active phase of the $\gamma$-ray source 1LHAASO J1219+2915. We propose that the detection of VHE $\gamma$-rays from NGC 4278 may be attributed to the presence of an active nucleus in its center. To reproduce the spectral energy distribution (SED) of NGC 4278, we employ a one-zone leptonic model, typically used for fitting broadband SEDs of BL Lacs, and find that a smaller magnetic field strength is required than that of typical TeV BL Lacs. Furthermore, NGC 4278 exhibits significantly lower luminosity in both radio and TeV bands when compared with typical TeV BL Lacs. In the radio-luminosity vs. Eddington-ratio plane, NGC 4278 shows greater similarity to Seyfert galaxies and LINERs rather than BL Lacs; however, it still roughly follows the extension towards lower luminosity seen in BL Lacs.

\end{abstract}


\section{Introduction} 

Radio-loud active galactic nuclei (RL-AGNs) account for about $15\%-20\%$ of the AGN population (\citealt{1995PASP..107..803U}) and are characterized by powerful relativistic jets. Blazars, a special type of RL-AGNs, have their jets pointing nearly along the line of sight and are divided into BL Lacertae objects (BL Lacs) and flat spectrum radio quasars (FSRQs) based on their emission line features in the optical band (\citealt{1997A&A...325..109S}). Blazars are the main extragalactic gigaelectronvolt-teraelectronvolt (GeV--TeV) $\gamma$-ray emitting objects, especially in the TeV band. In recent years, many other types of RL-AGNs have been confirmed as $\gamma$-ray emitters. For example, radio galaxies (\citealt{2020ApJS..247...33A}), Narrow-line Seyfert 1 galaxies (NLS1s, \citealt{2009ApJ...707L.142A}; \citealt{2015ApJ...798...43S}; \citealt{2019Galax...7...87D}; \citealt{2019JApA...40...39P}), compact steep-spectrum sources (CSSs, \citealt{2020ApJS..247...33A}; \citealt{2020ApJ...899....2Z}; \citealt{2022ApJ...927..221G}), compact symmetric objects (CSOs, \citealt{2016agnt.confE..60M}; \citealt{2020A&A...635A.185P}; \citealt{2020ApJ...899..141L}; \citealt{2021RAA....21..201G, 2022ApJ...939...78G, 2024RAA....24b5018G}), and low-luminosity FR0 type radio galaxies (\citealt{2016MNRAS.457....2G}; \citealt{2022RAA....22c5005F}). The large-scale extended regions of jets in AGNs also have been confirmed to emit detectable $\gamma$-rays (\citealt{2010Sci...328..725A}; \citealt{2016ApJ...826....1A}; \citealt{2024ApJ...965..163Y}). It is worth noting that in the TeV band, apart from several FSRQs and radio galaxies, BL Lacs are the main type of detected extragalactic $\gamma$-ray emitting AGNs\footnote{\url{http://tevcat2.uchicago.edu/}}. 

Recently, the Large High Altitude Air Shower Observatory (LHAASO) collaboration reported their first catalog (\citealt{2024ApJS..271...25C}). One new extragalactic TeV $\gamma$-ray source, 1LHAASO J1219+2915, is included. This source, located at high Galactic latitude, was detected with a significance level of 8.8 $\sigma$ during the active period and suggested to be associated with the AGN NGC 4278 (\citealt{2024ApJS..271...25C, 2024arXiv240507691C}). From a dedicated re-analysis of the LHAASO data, the distance estimated between NGC 4278 and 1LHAASO J1219+2915 is 0.03$\degr$ (\citealt{2024arXiv240507691C}).

NGC 4278, located at a redshift of $z=0.00216$ \citep{2009A&A...506.1107G}, has previously been classified as a low-ionization nuclear emission line region (LINER, \citealt{1980A&A....87..152H}). No AGNs of this type have been detected in the $\gamma$-ray band so far. NGC 4278 belongs to a RL-AGN category. However, the total radio luminosity of NGC 4278 is at least two orders of magnitude lower than that typically observed in RL-AGNs (\citealt{1998AJ....116.2682C}). The Very Long Baseline Array (VLBA) has resolved its two-sided symmetric steep-spectrum jets on a sub-parsec scale emerging from a flat-spectrum core region (\citealt{2005ApJ...622..178G}). Its radio flux density at 6 cm between 1972 and 2003 exhibits significant variability ($\geq$100\%) on timescale of years, and the outburst around 1985 may be related to the presence of an active nucleus (\citealt{2005ApJ...622..178G}). In the UV band, its nucleus is barely resolved and exhibits evident variability. Its UV flux increases by a factor of 1.6 during a period of 6 months, and the observed UV flare features, amplitude and scale time, are prominently similar to that observed in other low-luminosity AGNs (LLAGNs, \citealt{2009A&A...508..641C}). Using the Space Telescope Imaging Spectrograph, \cite{2009A&A...508..641C} obtained spectra of the nucleus, measured the broad components of the emission lines, and estimated the mass of the central supermassive black hole (SMBH) to be in the range of $7\times10^7-2\times10^9~M_{\bigodot}$. \cite{2010A&A...517A..33Y} reported that no evident short time-scale (hours and days) variability is observed in the nucleus of NGC 4278 during the six {\it Chandra} observations performed from 2005 February 3 to 2007 April 20. However, an increase in flux by a factor of $10\%$ is observed over a period of 1 hour during the {\it XMM-Newton} observation performed on 2004 May 23. Moreover, the nucleus flux in NGC 4278 increases by a factor of 3 over a few months and by a factor of 5 between the faintest and brightest observations separated by $\sim$3 years in X-rays. Additionally, NGC 4278 hosts numerous resolved low-mass X-ray binaries (LMXBs), more than one hundred of which exhibit long-term variability (\citealt{2009ApJS..181..605B}). In $\gamma$-rays, apart from a transient source, namely 1FLT J1219+2907 (with a photon spectral index of $3.3\pm0.4$), reported by the {\it Fermi}-LAT transient sources catalog with its position consistent with NGC 4278 from 2009 March 5 to 2009 April 5 (\citealt{2021ApJS..256...13B}), no other significant $\gamma$-ray activity has been detected from NGC 4278. Therefore, the origin of the detected $\gamma$-ray emission from 1LHAASO J1219+2915 and its association with NGC 4278 are still debated.

In this paper, we aim to study the radiation physics and the origin of very high energy (VHE) $\gamma$-rays from NGC 4278 by analysing its multiwavelength data. Data analysis of {\it Fermi}-LAT, {\it Chandra}, and {\it Swift}-XRT is presented in Section \ref{section2: DATA REDUCTION}. In Section \ref{section3: result}, we describe the X-ray emission property of the nucleus in NGC 4278. The construction and modeling of the broadband spectral energy distribution (SED) for the nucleus of NGC 4278 are presented in Section \ref{section4: SED Modeling}. Discussion on the origin of VHE $\gamma$-rays and the connection to the AGN NGC 4278 is given in Section \ref{section5: discussion}, while a summary is provided in Section \ref{section6: summary}. Throughout, the cosmological parameters $H_0=70$ km s$^{-1}$ Mpc$^{-1}$, $\Omega_{\rm m}=0.3$, and $\Omega_{\Lambda}=0.7$ are adopted.

\section{Observations and Data Reduction} \label{section2: DATA REDUCTION}

\subsection{{\it Fermi}-LAT}

We investigate the latest {\it Fermi}-LAT 14 yr Source Catalog (4FGL-DR4, \citealt{2022ApJS..260...53A}; \citealt{2023arXiv230712546B}) and find only two $\gamma$-ray sources located within a 1$\degr$ radius circle centered on the radio position of NGC 4278, namely 4FGL J1221.3+3010 and 4FGL J1217.9+3007. They are 0.929$\degr$ and 0.964$\degr$ distance from NGC 4278, and are associated with the BL Lacs 1ES 1218+304 and B2 1215+30, respectively. Both of these two BL Lacs have been detected in the TeV band and have been confirmed as TeV $\gamma$-ray emitters (\citealt{2006ApJ...642L.119A}; \citealt{2009ApJ...695.1370A}; \citealt{2012A&A...544A.142A}; \citealt{2013ApJ...779...92A}).

To further investigate if there are $\gamma$-ray emission from {\it Fermi}-LAT associated with NGC 4278, the Pass 8 data, covering $\sim$16 yrs from 2008 August 4 to 2024 April 24 (MJD 54682--60424), are extracted from Fermi Science Support Center\footnote{\url{https://fermi.gsfc.nasa.gov/ssc/data/access/}} for our analysis. We select the region of interest (ROI) centered at the radio position of NGC 4278 (R.A.=185.028$\degr$, Decl.=29.281$\degr$) with a radius of 15$\degr$. The publicly available software \texttt{Fermitools} (version 2.2.0) and \texttt{Fermipy} (v.1.2.2)\citep{Wood2017} are used for our binned maximum-likelihood analysis. We use event class ``SOURCE'' (evclass=128) and event type ``FRONT+BACK'' (evtype=3) for data analysis based on LAT data selection recommendations\footnote{\url{https://fermi.gsfc.nasa.gov/ssc/data/analysis/documentation/Cicerone/Cicerone_Data_Exploration/Data_preparation.html}}. The $\gamma$-ray events in the energy range of 0.1--300 GeV are selected with a standard data quality selection criteria ``(DATA\_QUAL\textgreater0)\&\&(LAT\_CONFIG==1)”. The instrument response function ``P8R3\_SOURCE\_V3”(\citealt{2018arXiv181011394B}) is used. A maximum zenith angle of $90\degr$ is set to reduce the  $\gamma$-ray contamination from the Earth limb. The background model in the ROI includes the isotropic background model iso\_P8R3\_V3\_v1.txt and the diffuse Galactic interstellar emission with the parameterized model gll\_iem\_v07.fits, as well as all the $\gamma$-ray sources listed in the 4FGL-DR4. The normalization of the isotropic background emission and the diffuse Galactic interstellar emission, together with the normalization and spectral parameters of the $\gamma$-ray point sources within a 6.5$\degr$ radius centered on the radio position of NGC 4278, are left free, whereas the parameters of those point sources lying beyond $6.5^\circ$ are fixed to their 4FGL-DR4 values.
 
The maximum likelihood test statistic (TS) is used to quantify the significance of the detection of $\gamma$-ray sources, which is defined as $\rm TS=2(log\mathcal{L}_{\rm src}-log\mathcal{L}_{\rm null})$ (\citealt{1996ApJ...461..396M}), where $\rm{\mathcal{L}_{src}}$ and $\rm{\mathcal{L}_{null}}$ are the maximum likelihood values for the background with and without a source, respectively. We generate a $3^\circ\times3^\circ$ residual TS map centered on the radio position of NGC 4278 and find that the maximum TS value in the residual TS map is $\sim$13, while the maximum TS value within a $1^\circ$ radius around NGC 4278 is approximately 9.

We further extend the analysis by exclusively focusing on PSF3 events above 1 GeV to implement more rigorous event filtering and generate a $3^\circ\times3^\circ$ residual map centered on NGC 4278. The resulting maximum TS value is approximately 4. According to the recommendations from the Fermi collaboration group, LAT catalogs necessitate a minimum TS value of 25 for a source to be deemed significantly detected. Since the significance corresponding to this threshold does not exceed 4$\sigma$, it suggests that the observed signal detection is likely attributable to random fluctuations in the background (\citealt{2010ApJS..188..405A}). The findings suggest that there is no identification of any new $\gamma$-ray source and no detection of significant GeV $\gamma$-ray emission associated with NGC 4278.

In order to investigate the presence of any similar high-energy activity during the active period in the VHE band from 2021 August 23 to 2022 January 10 (MJD 59449--59589, \citealt{2024arXiv240507691C}), we exclusively analyze the LAT data within the energy range of 0.1-300 GeV for this specific time frame. Employing a binned maximum-likelihood analysis approach, we add a point source with a power-law spectral form at the radio position of NGC 4278; however, our findings yield only a TS value of approximately 2.3, indicating no significant detection in the 0.1--300 GeV band during an active period of VHE $\gamma$-ray emission. Therefore, the 2$\sigma$ upper limits of the $\gamma$-ray flux for NGC 4278 in the 0.1--300 GeV band are derived.

\subsection{{\it Chandra}} 

NGC 4278 has been observed by the {\it Chandra} observatory using the Advanced CCD Imaging Spectrometer S-array (ACIS-S) in nine epochs. The shortest and longest exposure times are 1.4 ks and 110.7 ks, as listed in Table \ref{tab:X-ray}. The first seven observations (Obs-ID: 398, 4741, 7077, 7078, 7079, 7080, 7081) have been analyzed by \cite{2010A&A...517A..33Y}, and they found that all of these observations suffered the pile-up effect, with a pile-up fraction of $>5\%$. The first {\it Chandra} snapshot observation (Obs-ID: 398) with the short exposure time cannot produce a useful spectrum after considering the pile-up effect (\citealt{2010A&A...517A..33Y}), and thus it is excluded.

In this paper, we only analyze the last two observation data from the {\it Chandra} X-ray Observatory, which are included in the Chandra Data Collection \dataset[doi: 10.25574/11269]{https://doi.org/10.25574/11269} and \dataset[doi: 10.25574/12124]{https://doi.org/10.25574/12124}, corresponding to obs-ID 11269 on 2010 March 15 and obs-ID 12124 on 2010 March 20. The data are reduced with \texttt{CIAO} (v.4.15) and \texttt{CALDB} (v.4.10.4), and then we generate the level-2 event file in a standard procedure. We firstly check the pile-up effect of the two observations by creating pile-up maps with a \texttt{CIAO} tool of \texttt{pileup\_map}\footnote{\url{https://cxc.harvard.edu/ciao/ahelp/pileup_map.html}}. No pixel has the count rate per frame greater than 0.1 (corresponding to a pile-up fraction of $>5\%$), indicating no pile-up effect in the last two observations. We then extract the source photons from a circle centered on the radio position of NGC 4278 with a radius of $7^{''}$. The background is taken from an annulus region with inner and outer radii of $10^{''}$ and $20^{''}$, respectively. The spectrum is grouped to have at least 15 counts per bin and the $\chi^2$ minimization technique is used for the spectral analysis. The spectral fitting is performed using \texttt{XSPEC} (v.12.13.0c). We use the same model adopted in \cite{2010A&A...517A..33Y} to fit the spectra in the energy range of 0.5--8.0 keV, i.e., a single power-law absorbed by Galactic and host galaxy plus a thermal component (\texttt{mekal} in \texttt{XSPEC}). The neutral hydrogen column density of Galactic ($N^{\rm H}_{\rm gal}$) is fixed as $2.07\times10^{20}$ cm$^{-2}$ (\citealt{2005A&A...440..775K}) while the value of host galaxy ($N^{\rm H}_{\rm int}$) is set to be free. The fitting parameters of the two spectra are given in Table \ref{tab:X-ray} and the results of the first seven observations reported in \cite{2010A&A...517A..33Y} are also presented in the table for comparison. The derived temperatures of the thermal component in the two spectra are almost identical to that reported in \cite{2010A&A...517A..33Y}. However, we can only obtain an upper limit of $N^{\rm H}_{\rm int}$ for the two observations.  

\subsection{{\it Swift}-XRT}

{\it Swift} X-Ray Telescope ({\it Swift}-XRT) has performed 4 observations to NGC 4278 in the Photon Counting readout mode between 2017 December and 2021 November. The analysis is based on only two sets of observational data \footnote{During the other two observations, the radio position of NGC 4278 falls outside the XRT's view field.}, specifically obtained on 2021 Feb 26 (MJD 59271) with an exposure time of 173 s and on 2021 Nov 28 (MJD 59546) with an exposure time of 923 s. It is worth noting that the observational data from 2021 Feb 26 (MJD 59271) have a minimal impact on the spectral fit due to the short exposure time. The result of spectral fit is primarily influenced by the observational data from 2021 Nov 28 (MJD 59546). The data are processed using the \texttt{XRTDAS} software package (v.3.7.0). The softwave package was developed by the ASI Space Science Data Center and released by the NASA High Energy Astrophysics Science Archive Research Center (HEASARC) in the \texttt{HEASoft} package (v.6.30.1). The calibration files from XRT \texttt{CALDB} (version 20220803) are used within \texttt{xrtpipeline} to calibrate and clean the events. Considering the short exposure time of the two observations, we merge the two XRT event files together using the \texttt{xselect} package and produce a combined average spectrum. The source photons are extracted from a circle centered on the radio position of NCG 4278 with a radius of 20 pixels ($\sim47^{\prime\prime}$). The background is taken from an annulus with an inner and outer radii of 30 pixels ($\sim71^{\prime\prime}$) and 45 pixels ($\sim106^{\prime\prime}$), respectively. We utilize the public software \texttt{XSPEC} (v.12.13.0c) to fit the unbinned spectrum and employ C-statistic minimization to assess the goodness of fit. The energy channels below 0.5 keV and above 8 keV are excluded during fitting to maintain a consistent energy range with that derived from {\it Chandra} spectrum fitting. Same as the data analysis of {\it Chandra}, the spectrum is fitted by a single power-law absorbed by two absorption components; the neutral hydrogen column density of Galactic is fixed at $N^{\rm H}_{\rm gal} = 2.07\times 10^{20}\rm cm^{-2}$ (\citealt{2005A&A...440..775K}) while the neutral hydrogen column density of host galaxy is fixed at $N^{\rm H}_{\rm int}= 2.43\times 10^{20}\rm cm^{-2}$, which represents an upper-limit value obtained through {\it Chandra} spectrum fitting (OBS-ID: 11269). We obtain a C-statistic value/d.o.f of 24.07/30 and calculate the corrected flux in the energy range of $0.5-8.0$ keV, as listed in Table \ref{tab:X-ray}.

\section{X-ray Emission of the nucleus in NGC 4278} \label{section3: result}

\begin{figure*}
   \centering
   \includegraphics[angle=0,width=1\textwidth]{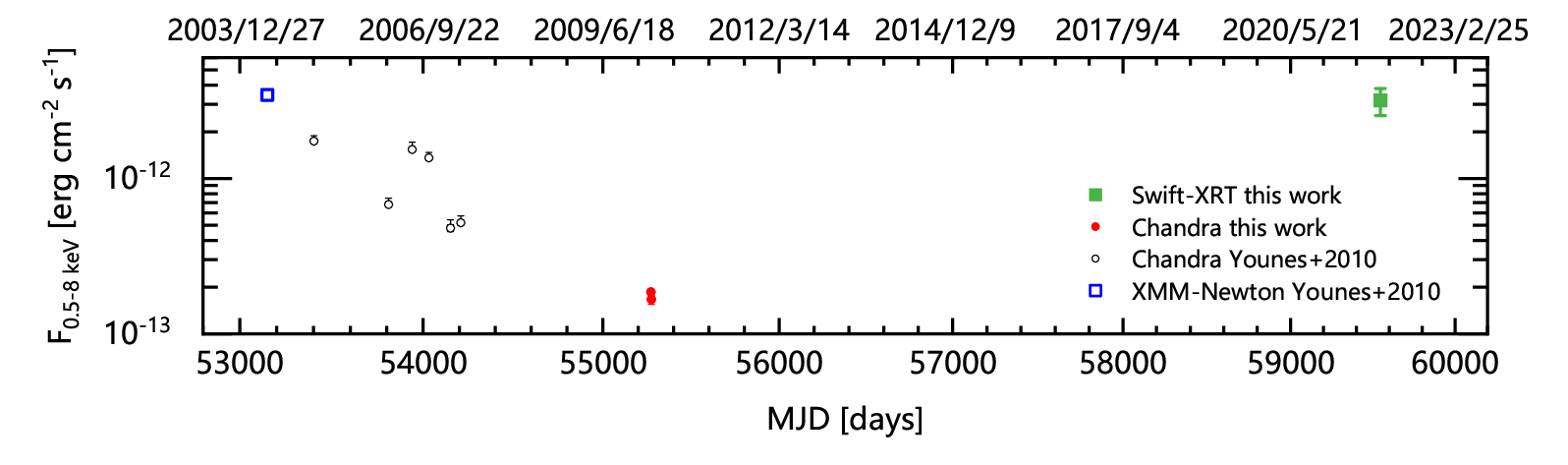}
   \caption{The X-ray light curve of NGC 4278 in the 0.5--8.0 keV band, obtained with the observations from {\it XMM-Newton}, {\it Chandra}, and {\it Swift}-XRT. The data points (open circles and square) before MJD 55000 are taken from \cite{2010A&A...517A..33Y}, while two red circles and one green square represent the analysis results conducted in this work.}\label{LC}
\end{figure*}

The X-ray light curve of NGC 4278 is compiled in Figure \ref{LC}, incorporating the archived X-ray data from \cite{2010A&A...517A..33Y} and the analysis results from the {\it Chandra} and {\it Swift}-XRT observations conducted in this study. Notably, NGC 4278 exhibits significant flux variations in the X-ray band. The {\it Swift}-XRT observations reveal an exceptionally high-flux state for NGC 4278 in the X-ray band, comparable to the historically highest flux obtained from the {\it XMM-Newton} observation on 2004 May 23. The ratio of the flux observed by {\it Swift}-XRT in 2021 to that observed by {\it Chandra} in 2010 exceeds one order of magnitude. Unfortunately, no additional data between 2010 to 2021 could be acquired to estimate variability timescales accurately. While previous studies have reported a $\sim$10\% increase in flux over a period of $\sim$1 hr during {\it XMM-Newton} observations (\citealt{2010A&A...517A..33Y}), due to limited exposure time during the {\it Swift}-XRT observations, further investigation into short timescale variability is not feasible in this study.

It should be noted that radio observations at 6 cm from 1972 to 2003 indicate variability in the flux density of NGC 4278, including an observed outburst around 1985, which may be associated with the presence of an active nucleus (\citealt{2005ApJ...622..178G}). The compact and symmetric two-sided radio morphology of NGC 4278 fulfills the criteria for classification as a CSO, and it is even included in a `bona fide' CSO sample \citep{2024ApJ...961..240K}. CSOs may be transient or episodic sources (\citealt{2021A&ARv..29....3O} for a review), with some $\gamma$-ray emitting CSOs reported to display episodic nuclear activity (\citealt{2020ApJ...899..141L}; \citealt{2021RAA....21..201G,2022ApJ...939...78G}). Recently, the LHAASO collaboration (\citealt{2024arXiv240507691C}) has reported that 1LHAASO J1219+2915 (associated with NGC 4278) exhibits an active period of VHE $\gamma$-ray emission in its 2.5-year light curve, specifically occurring between 2021 Aug 23 (MJD 59449) to 2022 Jan 10 (MJD 59589). It is noteworthy that the data point representing {\it Swift}-XRT observation in Figure \ref{LC} primarily corresponds to the X-ray flux of NGC 4278 on 2021 Nov 28 (MJD 59546). The observed high-flux state of NGC 4278 by {\it Swift}-XRT in 2021 could potentially be attributed to the reactivation of its nucleus, considering the previously low X-ray flux during the {\it Chandra} observations in 2010. This may be the reason that NGC 4278 has recently been detected at the VHE $\gamma$-ray band by LHAASO. 

\begin{figure*}
   \centering
   \includegraphics[angle=0, width=0.5\textwidth]{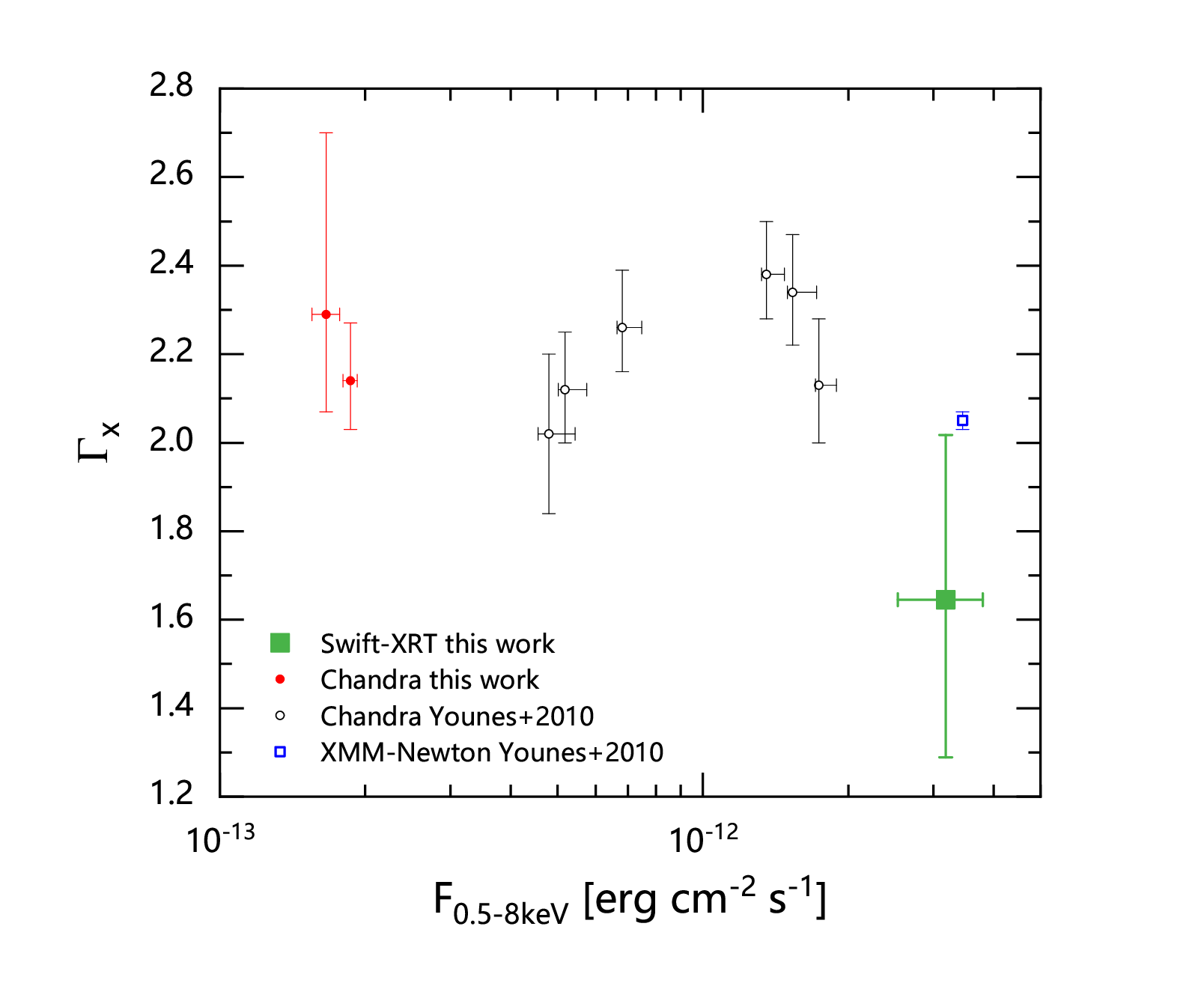}
   \caption{$\Gamma_{\rm X}$ vs. $F_{\rm 0.5-8~keV}$. The symbols are same as in Figure \ref{LC}.}\label{index}
\end{figure*}

The photon spectral index ($\Gamma_{\rm X}$) against the X-ray flux is depicted in Figure \ref{index}. Despite the presence of large errors in the data points, the spectral variation can still be observed. However, the X-rays of NGC 4278 do not exhibit the clear behavior of \emph{harder when brighter}, which is commonly observed among BL Lacs (e.g., \citealt{2009A&A...501..879T}; \citealt{2011ApJ...739...66T}; \citealt{2013ApJ...767....8Z}; \citealt{2022ApJ...938L...7D}; \citealt{2024ApJ...970L..22H}). 

\section{Constructing and Modeling the SED of Nucleus} \label{section4: SED Modeling}

\begin{figure*}
   \centering 
   \includegraphics[angle=0, width=0.6\textwidth]{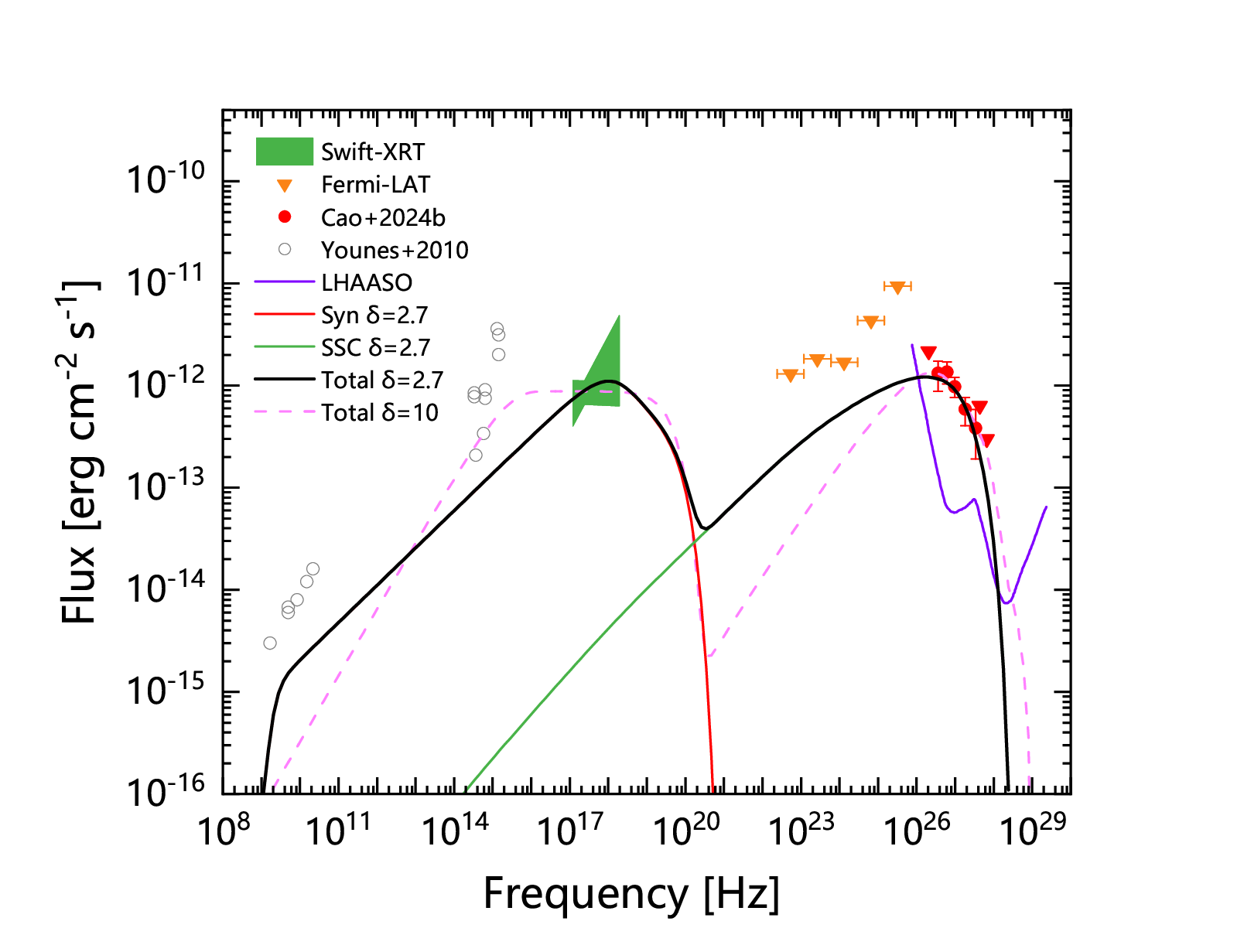}
\caption{Observed SEDs with model fitting for NGC 4278. The data in radio-optical-UV bands (gray open circles) are taken from \cite{2010A&A...517A..33Y}. The X-ray spectrum (green bowtie) is derived from the {\it Swift}-XRT in this work, while the orange inverted triangles indicate the $2\sigma$ upper limits of $\gamma$-ray flux observed by {\it Fermi}-LAT during the active period in the VHE band from 2021 August 23 to 2022 January 10 (MJD 59449--59589, \citealt{2024arXiv240507691C}). The TeV spectrum (red points and upper-limits) was obtained through LHAASO observations during an active phase of the source (\citealt{2024arXiv240507691C}). The black solid line represents the sum of each emission component, namely synchrotron radiation (red line) and SSC process (green line), for $\delta=2.7$, while the magenta dashed line indicates the total emission for $\delta=10$. The sensitivity curve of LHAASO is also presented.} \label{SED}
\end{figure*}

The reactivation of the nucleus, as mentioned above, likely results in the detection of VHE $\gamma$-rays and a high-flux state in X-rays from NGC 4278. Consequently, it is plausible that both the emission in X-rays and $\gamma$-rays originate from the same region. In order to further investigate the VHE $\gamma$-ray emission properties of NGC 4278, we collect its multiwavelength observation data from literature and construct a broadband SED. However, only a few non-simultaneous observations from \cite{2010A&A...517A..33Y} are available in the radio-optical-UV bands, as shown in Figure \ref{SED}. The sensitivity curves of LHAASO (\citealt{2022ChPhC..46c0003W}) and {\it Fermi}-LAT (\citealt{2013APh....43..348F}) are also presented in Figure \ref{SED}. The hard X-ray spectrum and soft TeV $\gamma$-ray spectrum should correspond to two distinct radiation components, resembling those typical TeV-emitting BL Lacs that have a high synchrotron radiation peak (e.g., \citealt{2012ApJ...752..157Z}). We employ a simple one-zone leptonic model, widely used in modeling the broadband SEDs of TeV BL Lacs (\citealt{2010MNRAS.401.1570T}; \citealt{2012ApJ...752..157Z}; \citealt{2014MNRAS.439.2933Y}), to reproduce the SED of NGC 4278, only considering synchrotron and synchrotron-self-Compton (SSC) processes of the relativistic electrons. 

The electron distribution is taken as a broken power law,

\begin{equation}
N(\gamma )= N_{0}\left\{ \begin{array}{ll}
\gamma ^{-p_1}  &  \mbox{ $\gamma_{\rm min}\leq\gamma \leq \gamma _{\rm b}$}, \\
\gamma _{\rm b}^{p_2-p_1} \gamma ^{-p_2}  &  \mbox{ $\gamma _{\rm b} <\gamma < \gamma _{\rm max} $.}
\end{array}
\right.
\end{equation}

The emission region is assumed as a sphere with radius $R$, magnetic field $B$, and a Doppler boosting factor $\delta$, where the $R$ value is fixed at $10^{16}$ cm, which is a typical value in SED modeling of TeV BL Lacs (\citealt{2010MNRAS.401.1570T}; \citealt{2012ApJ...752..157Z}). The VLBA observations indicate that NGC 4278 has mildly relativistic jets ($\beta\sim0.75$) with a viewing angle ($\theta$) of $2\degr\sim4\degr$, which result in a Doppler booting factor of 2.7 (\citealt{2005ApJ...622..178G}). Thus, we first fix $\delta$ to be 2.7. The values of $p_1$ and $p_2$ are constrained by the spectral indices at X-ray and TeV bands. Since $\gamma_{\min}$ cannot be constrained and is fixed as $\gamma_{\min}=1$, while $\gamma_{\max}$ is poorly constrained by the last observation point at TeV band. We adjust the parameters $B$, $N_0$, and $\gamma_{\rm b}$ to represent the observed SED of NGC 4278. The Klein--Nishina effect and the absorption of extragalactic background light (EBL; \citealt{2022ApJ...941...33F}) are also taken into account in the SED modeling. The SED fitting result is presented in Figure \ref{SED}. We obtain $B=7$ mG, $\gamma_{\rm b}=4\times10^6$, and $N_0=4.5\times10^4$ cm$^{-3}$ (see also in table \ref{tab:paras}). The derived value of $B=7$ mG is consistent with the estimated value of $B\ge5$ mG obtained by assuming that the electrons are cooling through the inverse Compton process on monthly timescale (\citealt{2024arXiv240507691C}).

Considering that the Doppler factor estimated from the radio data may underestimate the one in the $\gamma$-ray emission region of NGC 4278, we also adopt a fixed value of $\delta=10$ during the SED modeling. This choice is based on previous SED modeling studies for TeV BL Lacs (\citealt{2010MNRAS.401.1570T}; \citealt{2012ApJ...752..157Z}), where $\delta=10$ almost represents a lower-limit value of these TeV BL Lacs. In this case, a smaller magnetic field strength of $B=1$ mG is obtained. It should be noted that the fitting parameters $B$ and $\delta$ are degenerate (\citealt{2012ApJ...752..157Z}). If a high magnetic field strength is required to represent the broadband SED of NGC 4278, then a smaller $\delta$. For example, $\delta=1$ would be necessary with $B=0.03$ G.

Using the data from the LHAASO first catalog (\citealt{2024ApJS..271...25C}), \cite{2024ApJS..271...10W} also considered the one-zone synchrotron+SSC model to explain the X-ray and TeV $\gamma$-ray emission of NGC 4278. They obtained $B=0.04$ G by assuming $\theta=1.8\degr$ and a bulk Lorentz factor of $\Gamma=5$, while for $\theta=30\degr$ and $\Gamma=3$, they obtained $B=0.02$ G. It should be noted that their work adopted a significantly smaller value of $R\sim10^{14}$ cm, compared to ours and other BL Lacs in their study. Recently, \cite{2024arXiv240515657D} utilized the identical dataset and model to fit NGC 4278's SED as we did, reporting $B=8$ mG with $\delta=2.7$ and $R=6\times10^{15}$ cm, which aligns consistently with our findings.

As discussed above, the fitting parameters cannot be fully constrained due to the limited observational data of NGC 4278. We only obtain a set of model parameters that can provide an acceptable fit. Based on these SED fitting parameters, we also calculate the jet power ($P^{e^{\pm}}_{\rm jet}$) in case of the e$^{\pm}$ pair jet for NGC 4278 by assuming that the jet consist of electrons ($P_{\rm e}$), magnetic fields ($P_B$), and radiation ($P_{\rm r}$). In the case of $\delta=2.7$, the corresponding values are $P_{\rm e}=1.84\times10^{43}$ erg s$^{-1}$, $P_B=2.68\times 10^{38}$ erg s$^{-1}$, and $P_{\rm r}=1.69\times10^{41}$ erg s$^{-1}$, respectively. On the other hand, when $\delta=10$, the corresponding values are $P_{\rm e}=6.05\times10^{42}$ erg s$^{-1}$, $P_B=1.08\times10^{38}$ erg s$^{-1}$, and $P_{\rm r}=1.10\times10^{40}$ erg s$^{-1}$. These results are given in Table \ref{tab:paras}. 

\section{Discussion} \label{section5: discussion}

The angular resolution of LHAASO can achieve $<1\degr$ around 10 TeV (\citealt{2022ChPhC..46c0001M}). As stated in Section 2.1, there are two other GeV--TeV BL Lacs located within a distance of less than $1\degr$ from NGC 4278. Moreover, NGC 4278 harbors a substantial number of LMXBs (\citealt{2009ApJS..181..605B}; \citealt{2014ApJ...783...19D}), and $\gamma$-rays have been detected from several LMXBs in the Galaxy. We are wondering whether the VHE $\gamma$-rays detected by LHAASO from 1LHAASO J1219+2915 could potentially originate from either the two $\gamma$-ray emitting BL Lacs or these LMXBs in NGC 4278.

\subsection{Two TeV BL Lacs Located near NGC 4278}

The GeV--TeV emitting BL Lacs 1ES 1218+304 and B2 1215+30 are situated at a distance of 0.929$\degr$ and 0.964$\degr$, respectively, from the radio position of NGC 4278. Firstly, we conduct a literature search to identify the highest observed flux at the TeV band for the two BL Lacs. We then extrapolate their intrinsic $\gamma$-ray spectrum up to 100 TeV assuming a power-law spectral form. Subsequently, we calculate the observable spectral shape by taking into account the absorption of EBL and compare it with that of 1LHAASO J1219+2915. Two EBL models are considered in this analysis, corresponding to the strongest absorption (\citealt{2017A&A...603A..34F}) and weakest absorption (\citealt{2010cosp...38.2365K}), respectively.

\begin{figure*}
    \centering
    \includegraphics[angle=0,width=0.32\textwidth]{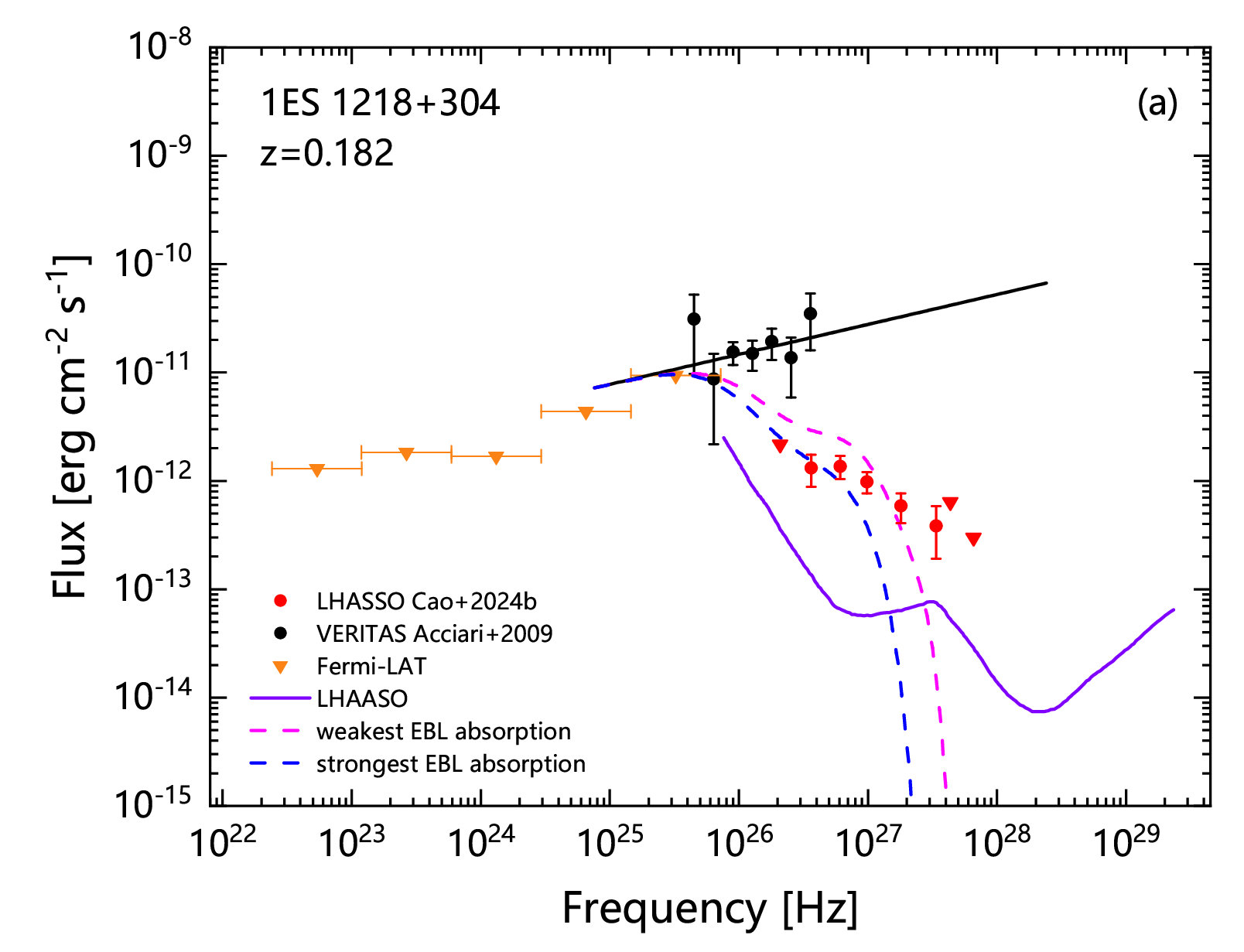}
    \includegraphics[angle=0,width=0.32\textwidth]{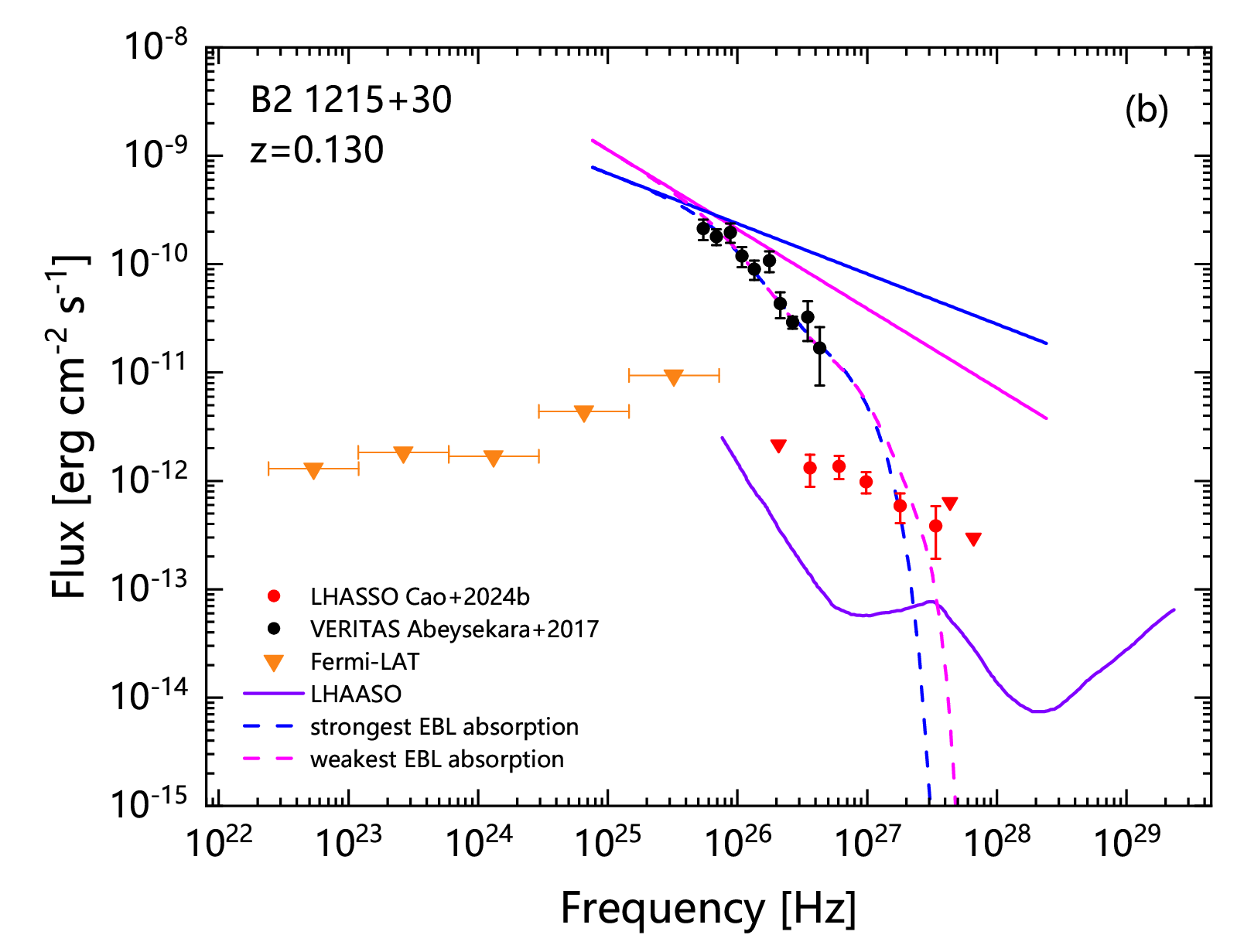}
    \includegraphics[angle=0,width=0.32\textwidth]{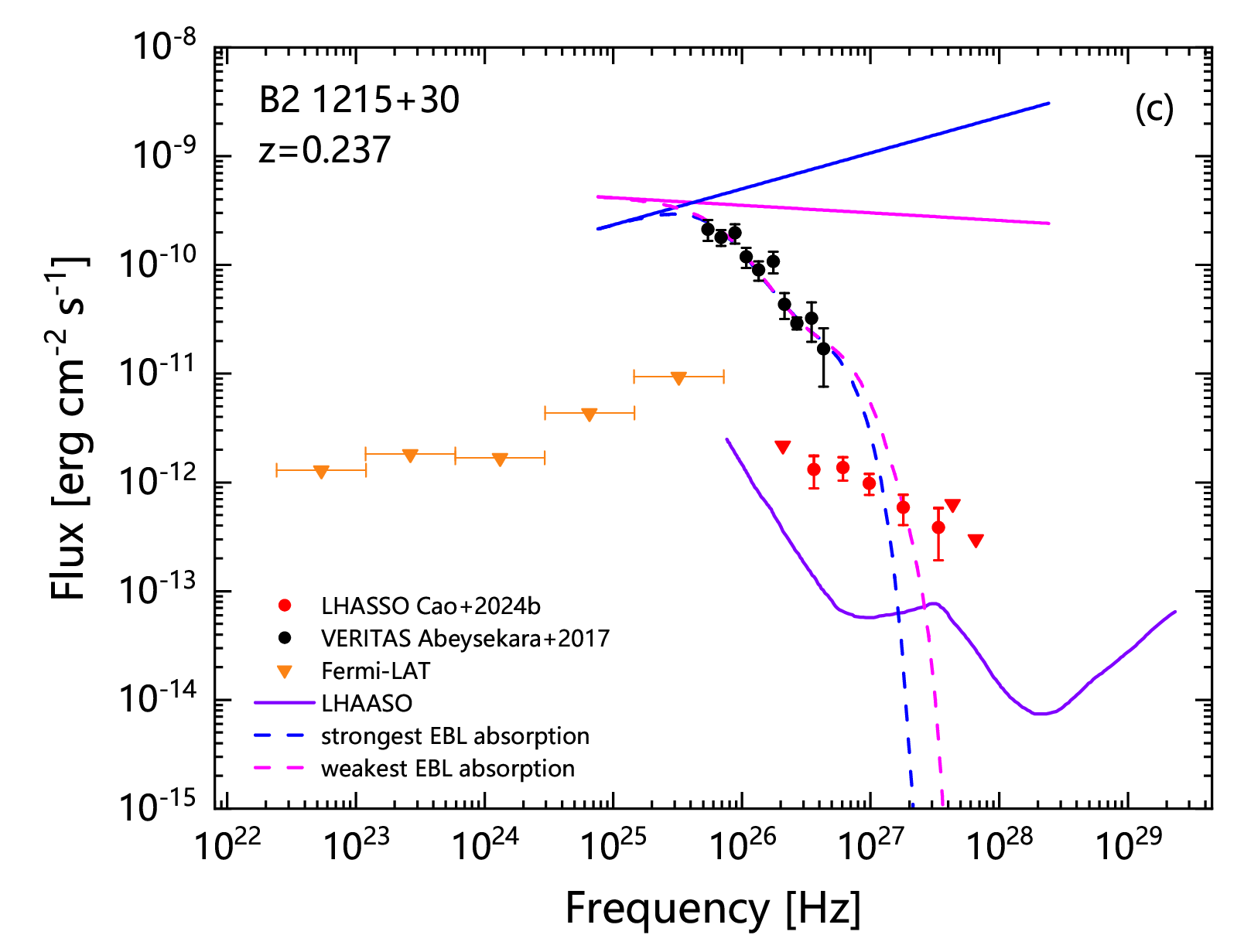}
    \caption{Panel (a): The intrinsic fluxes (black circles) of 1ES 1218+304 in the TeV band are taken from \cite{2009ApJ...695.1370A}. They are fitted with a power-law function and extrapolated up to the 100 TeV band (black solid line). Panel (b): The observed fluxes of B2 1215+30 in the TeV band are taken from \cite{2017ApJ...836..205A}. Considering its redshift of $z=0.130$, correcting into the intrinsic fluxes with the strongest and weakest models of EBL, fitting them with a power-law function, and extrapolating the fitting lines up to the 100 TeV band, we obtain the blue and magenta solid lines. Panel (c): Same as in Panel (b), but with the redshift of $z=0.237$. The orange inverted triangles (the $2\sigma$ upper limits in the GeV band) and the red symbols (the VHE emission during an active phase, \cite{2024arXiv240507691C}) are same as in Figure \ref{SED}. The sensitivity curve of LHAASO is also presented. The blue and magenta dashed lines represent the observable flux level, taking into account the absorption from both the strongest and weakest models of EBL, respectively. }
    \label{Two_BL}
\end{figure*}

For 1ES 1218+304, located at a redshift of $z=0.182$ (\citealt{2003A&A...412..399V}), the intrinsic $\gamma$-ray spectrum in the TeV band is taken from \cite{2009ApJ...695.1370A}. We extrapolate this intrinsic $\gamma$-ray spectrum to higher energy bands by assuming a power-law spectral shape, and subsequently derive the observable spectrum in the TeV band by taking into account the absorption of EBL, as illustrated in Figure \ref{Two_BL}(a). For B2 1215+30, we obtain the observed spectrum in the TeV band from \cite{2017ApJ...836..205A}. Considering that there is uncertainty regarding its redshift, i.e., either $z = 0.130$ (\citealt{2003ApJS..148..275A}) or $z = 0.237$ (\citealt{1993ApJS...84..109L}), we consider both cases. Firstly, we correct for EBL absorption on the observation data using both strongest and weakest EBL models to obtain an intrinsic $\gamma$-ray spectrum and extrapolate this obtained intrinsic $\gamma$-ray spectrum up to 100 TeV (the blue and magenta solid lines in Figure \ref{Two_BL}) by assuming a power-law spectral form. Subsequently, considering again the effect of EBL absorption, we calculate the observable flux in the TeV band, represented by the blue and magenta dashed lines shown in Figures \ref{Two_BL}(b) and \ref{Two_BL}(c).

The LHAASO is capable of detecting both 1ES 1218+304 and B2 1215+30 when they are in a high-flux state, as depicted in Figure \ref{Two_BL}. However, the estimated observable spectra of the two BL Lacs fail to represent the observed highest energy photon of 1LHAASO J1219+2915 even considering the weakest EBL absorption model. In other words, due to their high redshift, it appears impossible for the LHAASO to detect $\gamma$-ray photons with energies exceeding 10 TeV from the two BL Lacs, even when considering the weakest EBL absorption model. Therefore, the VHE $\gamma$-ray emission from 1LHAASO J1219+2915 should not be contributed by the two BL Lacs. 

\subsection{LMXBs in NGC 4278}

\begin{figure*}
    \centering
    \includegraphics[angle=0,width=0.6\textwidth]{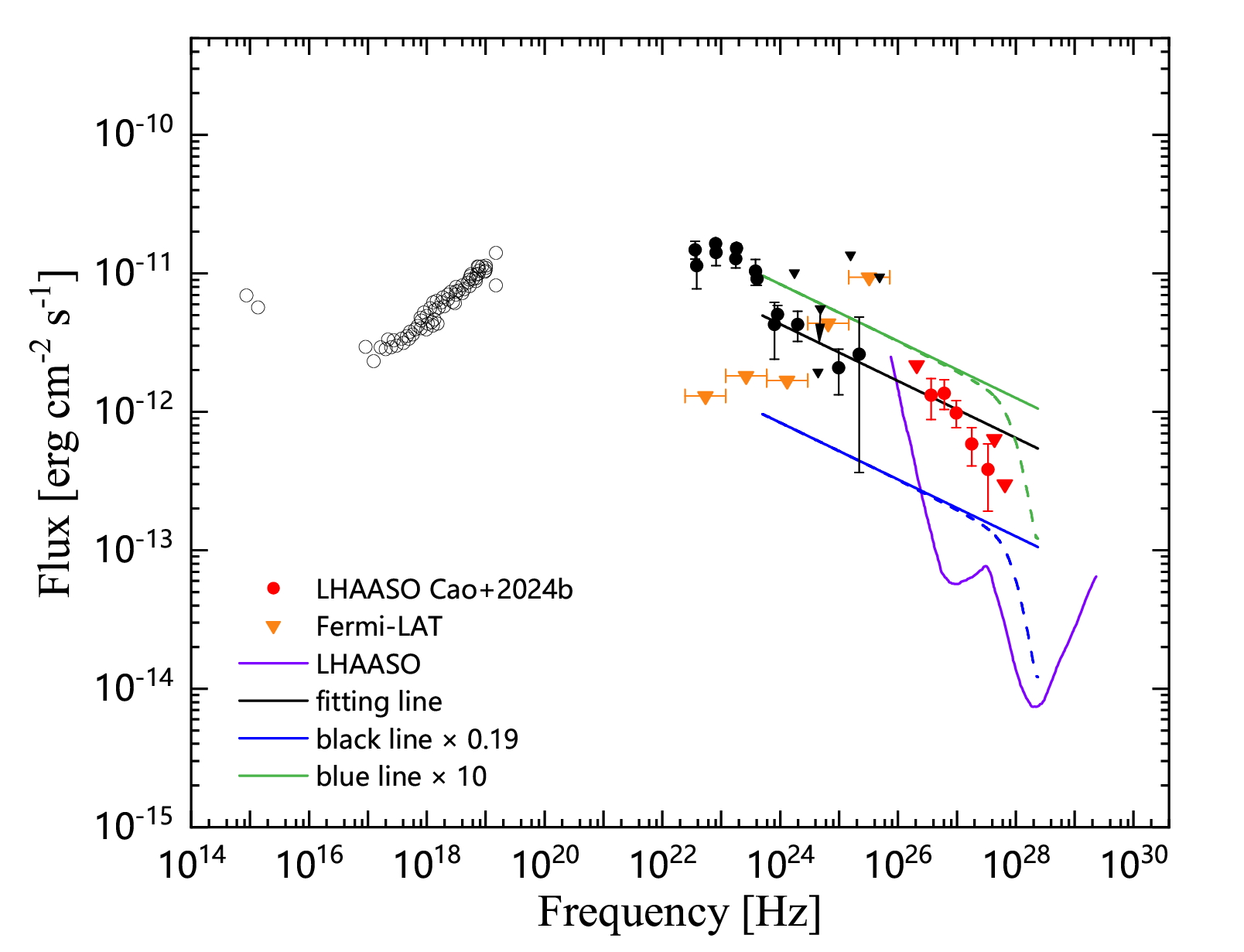}
    \caption{Schematic diagram illustrating the expected flux from the LMXBs in NGC 4278. The observed data (black circles and upper-limits) of PSR J1023+0038 are taken from \cite{2014ApJ...797..111L} and \cite{2018RAA....18..127X}. The orange inverted triangles (the $2\sigma$ upper limits in the GeV band) and the red symbols (the VHE emission during an active phase, \cite{2024arXiv240507691C}) are same as in Figure \ref{SED}. The sensitivity curve of LHAASO is also presented. The black solid line represents the power-law fitting line for the last four detection data points of PSR J1023+0038 at GeV band, and it is extrapolated to the 100 TeV band. The blue solid line indicates the possible flux level of the brightest LMXB in NGC 4278, which is obtained by multiplying the black solid line by a factor of 0.19. The green solid line is simply the solid blue line multiplied by 10, assuming that these are several $\gamma$-ray emitting LMXBs in NGC 4278. The blue and green dashed lines present the results after considering the EBL absorption. }
    \label{LMXB1023}
\end{figure*}

Two identified LMXBs (1SXPS J042749.2-670434 and PSR J1023+0038), along with seven associated LMXBs that are not definitively identified, have been reported in the 4FGL-DR4 (\citealt{2022ApJS..260...53A}; \citealt{2023arXiv230712546B}). Additionally, several other Galactic LMXBs are also suggested to be the $\gamma$-ray sources (e.g., \citealt{2015ApJ...808...17X}; \citealt{2022MNRAS.512.1141H}; \citealt{2022MNRAS.510.5187K}). In NGC 4278, there exist multiple LMXBs with an X-ray luminosity exceeding $10^{38}$~erg~s$^{-1}$ (\citealt{2009ApJS..181..605B}; \citealt{2010ApJ...725.1824F}), which are likely to be sources of $\gamma$-ray emission as well. Therefore, we estimate the potential $\gamma$-ray flux from these LMXBs in NGC 4278 by assuming that they exhibit a similar X-ray to $\gamma$-ray flux ratio as the Galactic $\gamma$-ray emitting LMXBs, i.e., 

\begin{equation}
    \frac{F_{\gamma,\rm LMXB}}{F_{\rm X,LMXB}}=\frac{F_{\gamma,\rm J1023}}{F_{\rm X,J1023}},
\end{equation}

where $F_{\gamma,\rm LMXB}$ and $F_{\rm X,LMXB}$ are the fluxes at $\gamma$-ray and X-ray bands of the LMXBs in NGC 4278, $F_{\gamma,\rm J1023}$ and
$F_{\rm X,J1023}$ are the $\gamma$-ray and X-ray fluxes of the LMXB PSR J1023+0038. PSR J1023+0038 is a well observed Galactic $\gamma$-ray emitting LMXB. The observational data for PSR J1023+0038 are obtained from \cite{2014ApJ...797..111L} and
\cite{2018RAA....18..127X}, as shown in Figure \ref{LMXB1023} (black symbols). Taking $F_{\rm X,J1023}=4.01\times10^{-13}\rm\ erg\ cm^{-2}\ s^{-1}$ in the $0.5-10$ keV band (\citealt{2016A&A...594A..31C}) and $F_{\rm X,LMXB}=7.79\times10^{-14}\rm\ erg\ cm^{-2}\ s^{-1}$ in the $0.3-8$ keV band (\citealt{2010ApJ...725.1824F}), where the maximum value of $F_{\rm X,LMXB}$ among these LMXBs in NGC 4278 is considered, the $\gamma$-ray fluxes of the brightest LMXB in NGC 4278 can be estimated as $F_{\gamma,\rm LMXB}=0.19F_{\gamma,\rm J1023}$. 

We extrapolate the GeV spectrum of PSR J1023+0038 up to the TeV band by assuming that it follows the same power-law distribution (the black solid line in Figure \ref{LMXB1023}). Multiplying the black solid line by 0.19, we obtain the possible $\gamma$-ray flux (the blue solid line) of the brightest LMXB in NGC 4278. The blue dashed line in Figure \ref{LMXB1023} presents the result after considering EBL absorption. It is evident that if there is only one $\gamma$-ray emitting LMXB in NGC 4278, its emission would not be detectable by {\it Fermi}-LAT but could potentially be detected by LHAASO. NGC 4278 hosts several LMXBs with X-ray luminosity exceeding $10^{38}$ erg s$^{-1}$, some of which exhibit significant variability (\citealt{2010ApJ...725.1824F}). Therefore, it is plausible that the observed $\gamma$-rays from 1LHAASO J1219+2915 can be attributed to some LMXBs in NGC 4278. However, no detection has been made for any LMXB at TeV energy band.

In addition, the spectrum at the TeV band of LMXB may not follow the same power-law spectral shape as that at the GeV band; for example, a flatter spectrum is observed in high-mass X-ray binaries (HMXBs) at the TeV band, such as HESS J1018–589 A/1FGL J1018.6–5856 (\citealt{2015A&A...577A.131H}). Considering the presence of multiple bright LMXBs in NGC 4278, we cannot dismiss the possibility that the detection of 1LHAASO J1219+2915 could be attributed to these LMXBs within NGC 4278.

\subsection{NGC 4278: a BL Lac or a LINER}

\begin{figure*}
    \centering
    \includegraphics[angle=0,width=0.5\textwidth]{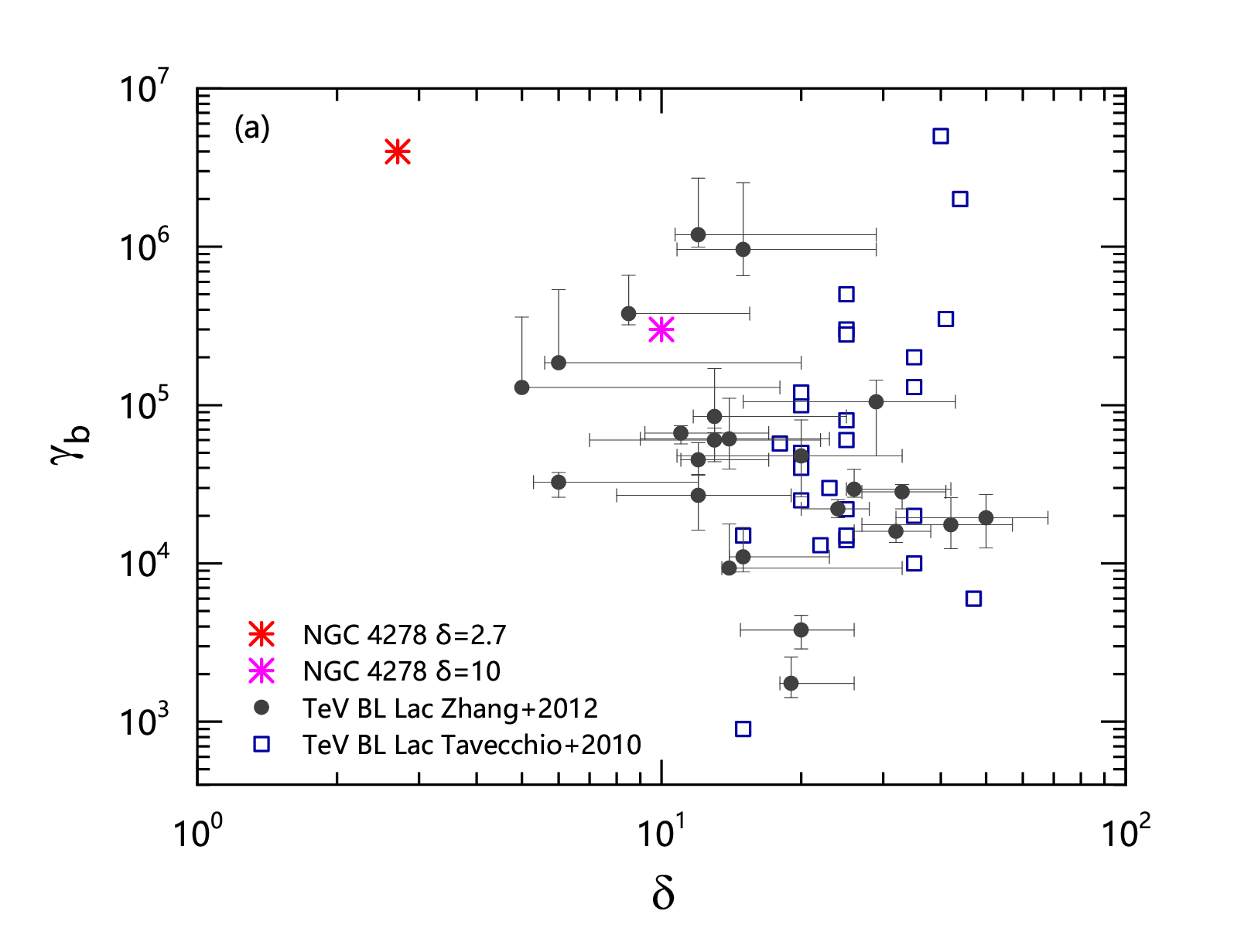}\hspace{-5mm}
    \includegraphics[angle=0,width=0.5\textwidth]{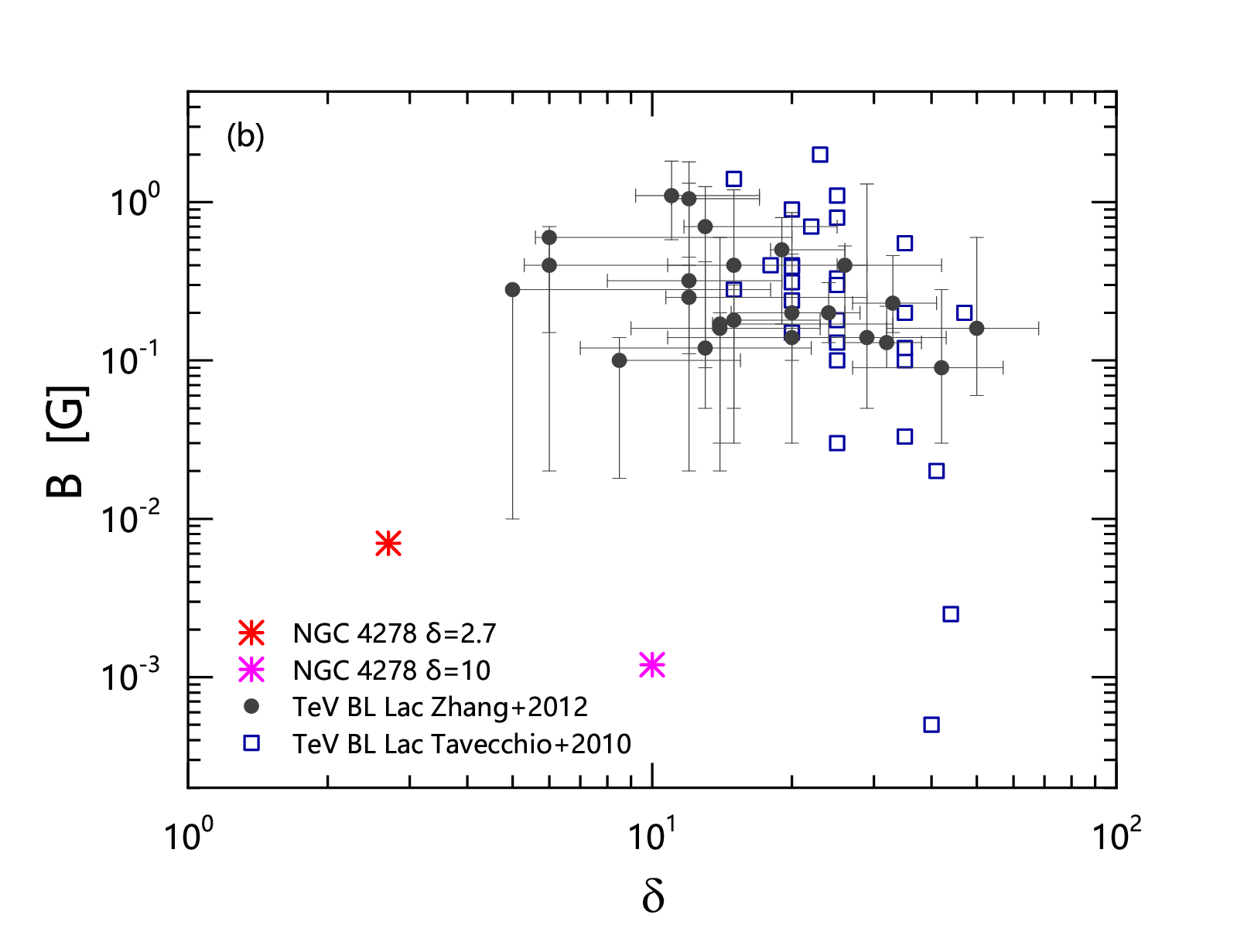}
    \caption{$\gamma_{\rm b}$ and $B$ as a function of $\delta$. The data of these TeV BL Lacs are taken from \cite{2012ApJ...752..157Z} and \cite{2010MNRAS.401.1570T}.}
    \label{gamb_B_delta}
\end{figure*}

\begin{figure*}
    \centering
   \includegraphics[angle=0,width=0.5\textwidth]{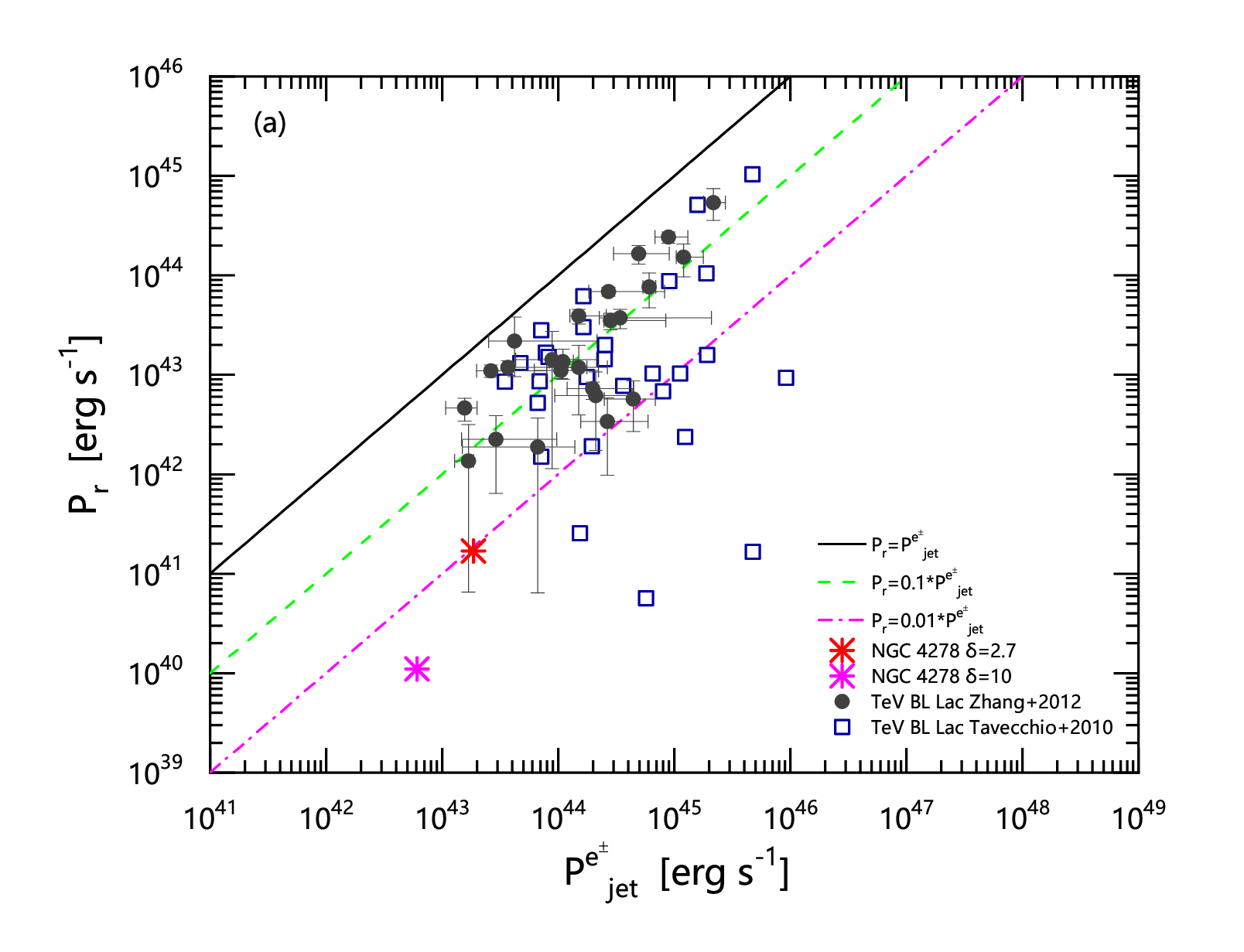}\hspace{-5mm}
   \includegraphics[angle=0,width=0.5\textwidth]{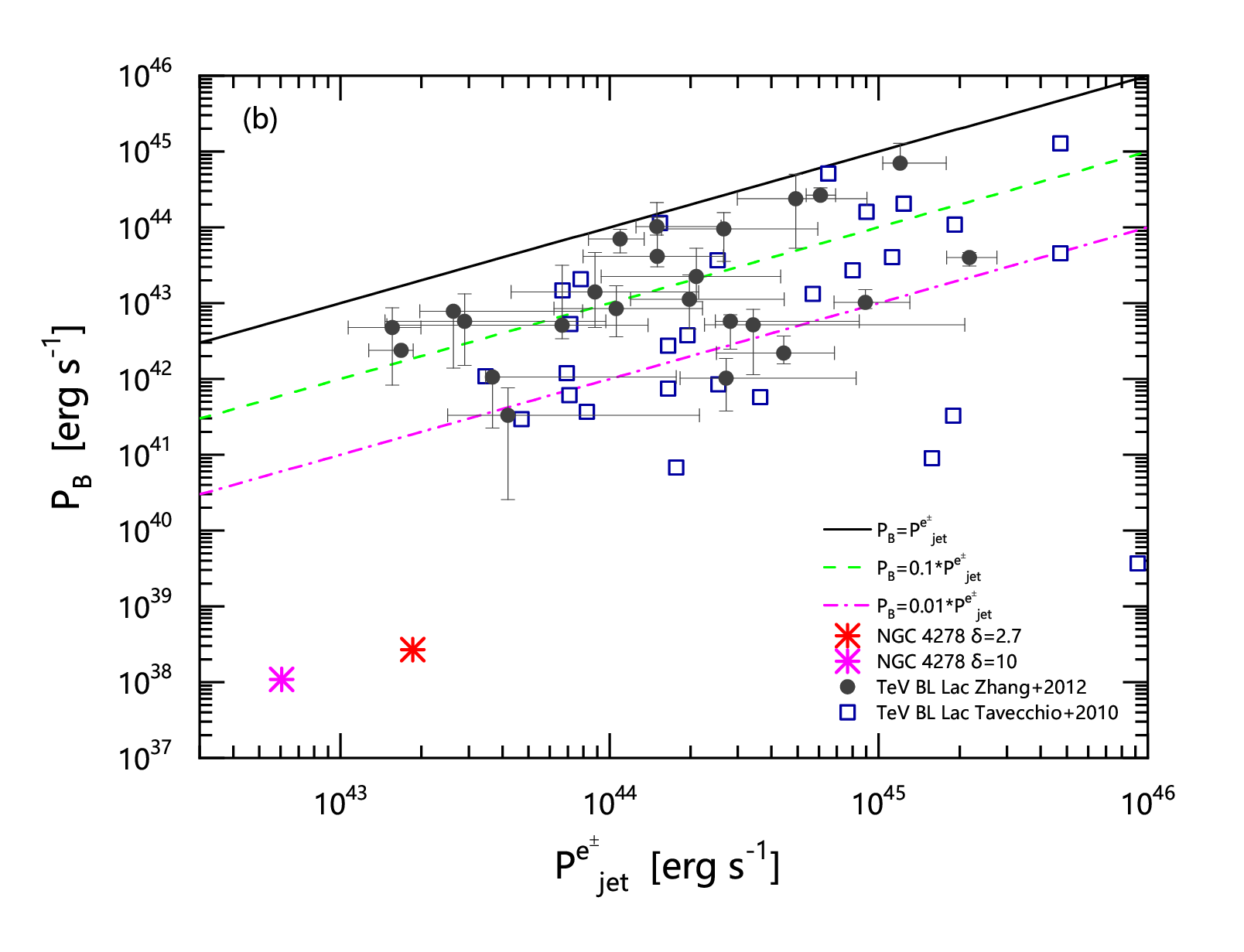}
    \caption{$P_{\rm r}$ and $P_{B}$ as a function of $P^{e^{\pm}}_{\rm jet}$. The data of these TeV BL Lacs are taken from \cite{2012ApJ...752..157Z} and \cite{2010MNRAS.401.1570T}.}
    \label{Pj_pr}
\end{figure*}

The compact radio morphology and flat radio spectrum of NGC 4278 suggest its similarity to BL Lacs (\citealt{2005ApJ...622..178G}). The SED of NGC 4278 can be represented by a one-zone synchrotron+SSC model, similar to the typical TeV BL Lacs. We compare the derived jet physical parameters of NGC 4278 with that of a TeV-BL Lac sample from \cite{2010MNRAS.401.1570T} and \cite{2012ApJ...752..157Z}. The characteristics of NGC 4278 appear to deviate from those typical TeV BL Lacs, as shown in Figure \ref{gamb_B_delta}. For $\delta$=10, the $\gamma_{\rm b}$ value of NGC 4278 is comparable to other BL Lacs, but a smaller $B$ value is required. Even when considering a significantly smaller value of $\delta$=2.7 compared to other BL Lacs, the derived $B$ value remains lower than that of most other BL Lacs. In the $P_{\rm jet}^{e^{\pm}}-P_{\rm r}$ plane (Figure \ref{Pj_pr}), NGC 4278 follows the sequence of typical TeV BL Lacs but resides at the low-power end. Nevertheless, when examining the $P_{\rm jet}^{e^{\pm}}-P_{B}$ plane, it becomes evident that NGC 4278 stands apart from these TeV BL Lacs due to an unusually smaller ratio of $P_B$ to $P_{\rm jet}^{e^{\pm}}$.

\begin{figure*}
    \centering
    \includegraphics[angle=0,width=0.5\textwidth]{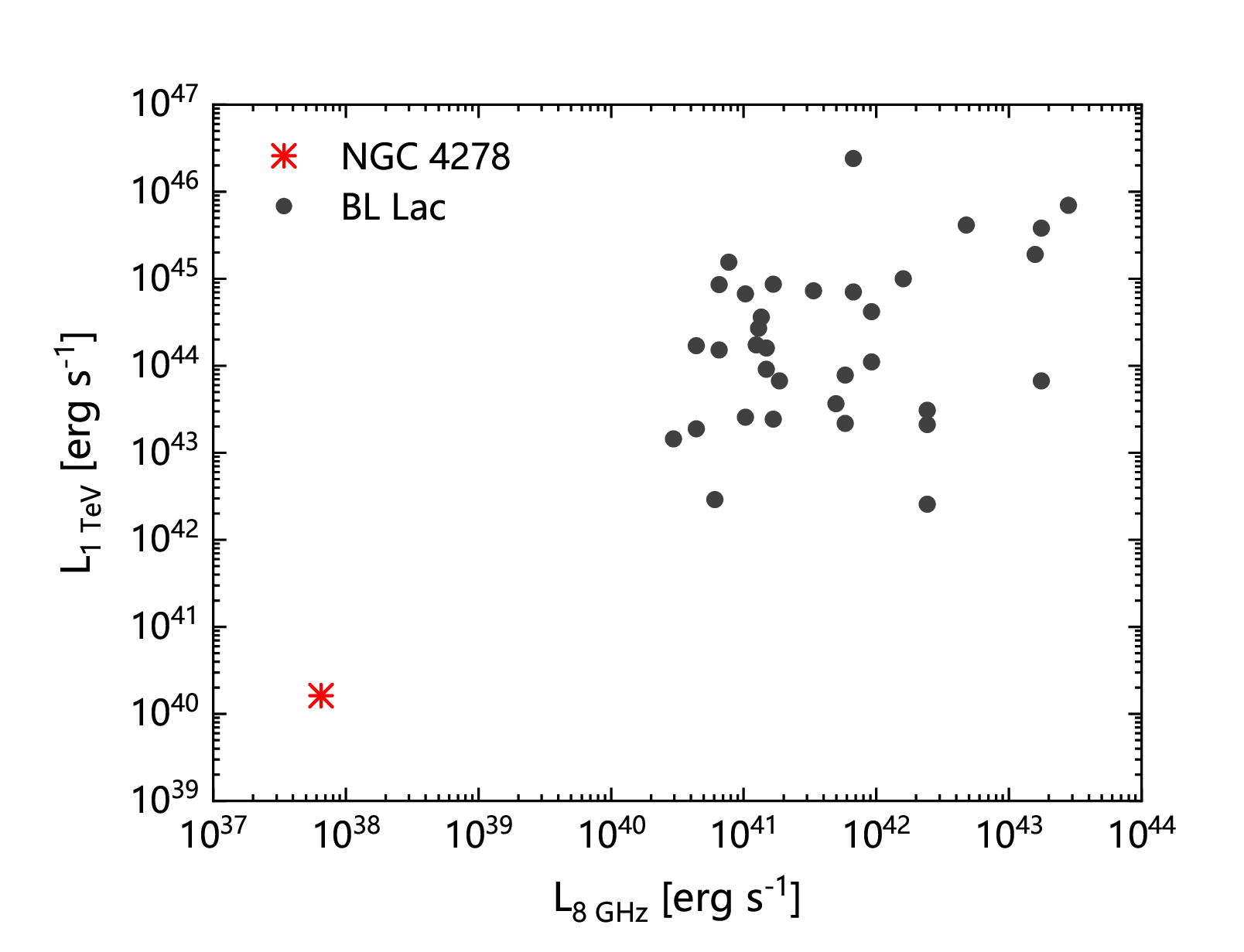}\hspace{-5mm}
    \includegraphics[angle=0,width=0.5\textwidth]{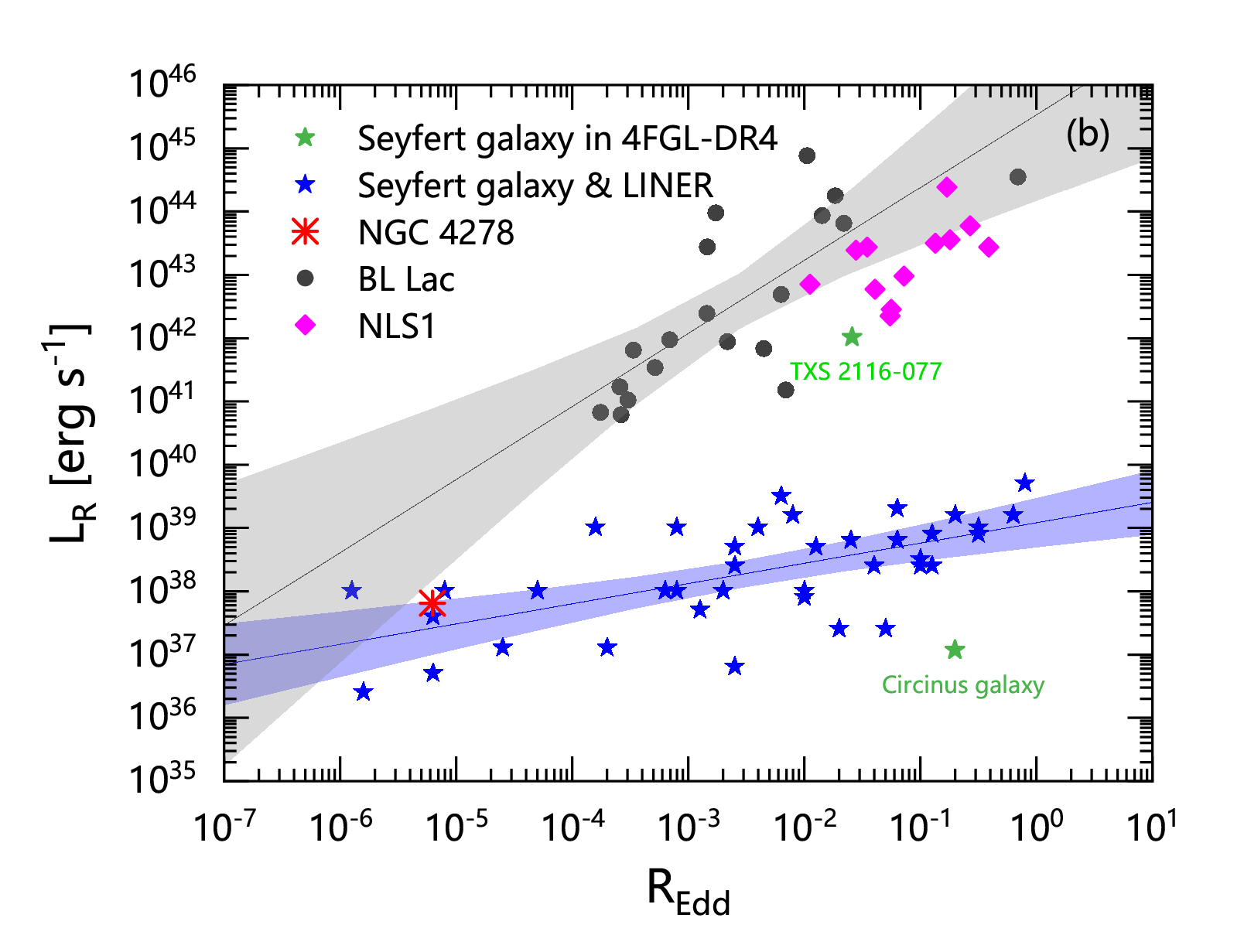}
    \caption{Panel (a): $L_{\rm 1~TeV}$ vs. $L_{\rm 8~GHz}$. The data of $L_{\rm 8~GHz}$ for both NGC 4278 and TeV BL Lacs are taken from the RFC, while the $L_{\rm 1~TeV}$ values of the TeV BL Lacs are from \cite{2012ApJ...752..157Z}. Panel (b): $L_{\rm R}$ vs. $R_{\rm Edd}$, where $L_{\rm R}$ is the luminosity at 8 GHz for NGC 4278, BL Lacs, NLS1s, and two Seyfert galaxies in the 4FGL-DR4 taken from the RFC, while it is the luminosity at 5 GHz for Seyfert galaxies and LINERs taken from \cite{2007ApJ...658..815S}. The values of $R_{\rm Edd}$ for Seyfert galaxies and LINERs are also taken from \cite{2007ApJ...658..815S}, while those for BL Lacs, NLS1s, and TXS 2116--077 are obtained from \cite{2020ApJ...899....2Z}, where $R_{\rm Edd}$ of Circinus galaxy is taken from \cite{2003ApJ...590..162G}. Note that the BL Lac sample in Panel (b) differs partially from that in Panel (a); for more details, refer to Section 5.3.}
    \label{LR-Redd}
\end{figure*}

The flux densities at 1 TeV of some typical BL Lacs are documented in \cite{2012ApJ...752..157Z}. We thus compile the radio luminosity at 8 GHz ($L_{\rm 8~GHz}$) of NGC 4278 and these BL Lacs from the Radio Fundamental Catalog (RFC)\footnote{\url{https://astrogeo.smce.nasa.gov/rfc/}}, and plot the $\gamma$-ray luminosity at 1 TeV ($L_{\rm 1~TeV}$) against $L_{\rm 8~GHz}$ in Figure \ref{LR-Redd}(a). No significant correlation between $L_{\rm 8~GHz}$ and $L_{\rm 1~TeV}$ for these typical TeV BL Lacs is observed. Moreover, NGC 4278 stands out as it is clearly separated from these typical BL Lacs and positioned at the lower end of the luminosity scale. The luminosity of NGC 4278 is more than two orders of magnitude lower than that of these TeV BL Lacs in both radio and $\gamma$-ray bands.

In Figure \ref{LR-Redd}(b), we plot the radio luminosity ($L_{\rm R}$) against the Eddington ratio ($R_{\rm Edd}$) for a sample of Seyfert galaxies and LINERs from \cite{2007ApJ...658..815S}\footnote{The only exclusion is NGC 1275 due to its classification as a TeV $\gamma$-ray emitting radio galaxy (\citealt{2012A&A...539L...2A}).}, along with a sample of BL Lacs and $\gamma$-ray emitting NLS1s from \cite{2020ApJ...899....2Z} and references therein, as well as two Seyfert galaxies\footnote{Only four associated Seyfert galaxies, which have not yet been definitively identified, are presented in the 4FGL-DR4. PKS 0000--160 and NVSS J01531--1844 lack available data and are not included in this study.} from the 4FGL-DR4 (\citealt{2022ApJS..260...53A}; \citealt{2023arXiv230712546B}). The values of $R_{\rm Edd}$ for the two Seyfert galaxies (TXS 2116--077 and Circinus galaxy) are taken from \cite{2020ApJ...899....2Z} and \cite{2003ApJ...590..162G}, respectively. Note that the BL Lac sample in Figure \ref{LR-Redd}(b) differs partially from that in Figure \ref{LR-Redd}(a). Not all BL Lacs shown in Figure \ref{LR-Redd}(b) are TeV sources, and not all BL Lacs depicted in Figure \ref{LR-Redd}(a) have available $R_{\rm Edd}$ data. However, there is an overlap between the two samples. For Seyfert galaxies and LINERs from \cite{2007ApJ...658..815S}, $L_{\rm R}$ represents the luminosity at 5 GHz, while for other sources it corresponds to $L_{\rm 8~GHz}$, where the $L_{\rm 8~GHz}$ values are also derived from the observation data in the RFC. 

In the $R_{\rm Edd}-L_{\rm R}$ plane, NGC 4278 is situated within the region occupied by Seyfert galaxies and LINERs, distinctly separated from these BL Lacs. Circinus galaxy also shares similarities with these Seyfert galaxies and LINERs, while TXS 2116--077 obviously belongs to $\gamma$-ray emitting NLS1s, as suggested in \cite{2019Galax...7...87D}. The BL Lac sample and Seyfert galaxy \& LINER sample both (neither including NGC 4278) exhibit a clear correlation between $L_{\rm R}$ and $R_{\rm Edd}$, whereas this correlation is absent in $\gamma$-ray emitting NLS1s (including TXS 2116--077). The Pearson correlation analysis yields a coefficient of $r=0.629$ and a chance probability of $p=2.34\times 10^{-5}$ for the Seyfert galaxy \& LINER sample, while obtaining $r=0.769$ with $p=1.17\times 10^{-4}$ for these BL Lacs, and deriving $r=0.596$ with $p=0.031$ for the $\gamma$-ray emitting NLS1s. The position of NGC 4278 slightly deviates from the low-luminosity extension of the 95\% confidence band of the best linear fit for BL Lacs. However, the Eddington ratio of NGC 4278 is more than one order of magnitude lower than that of BL Lacs, which may be the reason for its low luminosity. Considering all these aspects, NGC 4278 does not resemble a BL Lac; instead, it exhibits more similarities to Seyfert galaxies and LINERs.

\section{summary}\label{section6: summary}

NGC 4278 was classified as a LINER-type AGN and no associated $\gamma$-ray source counterpart reported in the 4FGL-DR4 (\citealt{2022ApJS..260...53A}; \citealt{2023arXiv230712546B}). By reanalyzing all the {\it Fermi}-LAT observation data from 2008 Aug 4 to 2024 Apr 24 around the radio position of NGC 4278, no significant $\gamma$-ray emission associated with NGC 4278 is found. Interestingly, NGC 4278 has been detected by LHAASO and is reported to be associated with 1LHAASO J1219+2915 at a significance level of 8.8 $\sigma$ (\citealt{2024arXiv240507691C}). To investigate the origin of VHE emission from this source, we examine its X-ray observations obtained by {\it Swift}-XRT, {\it Chandra}, and {\it XMM-Newton}, and find only one valid pointing (on 2021 Nov 28, MJD 59546) from {\it Swift}-XRT during the LHAASO observations. Moreover, this particular pointing was conducted during an active period (from August 23, 2021 to January 10, 2022) for the source 1LHAASO J1219+2915. Interestingly, the derived X-ray flux from {\it Swift}-XRT is more than one order of magnitude higher than the archive X-ray flux obtained by {\it Chandra} observation $\sim$11.7 years ago, indicating that the X-ray emission of NGC 4278 is in a high-flux state during the active phase of the source 1LHAASO J1219+2915. Together with the mildly relativistic jets detected in the radio band using the VLBA observations, we propose that the TeV $\gamma$-rays of 1LHAASO J1219+2915 originate from an atypical BL Lac-type AGN in the center of NGC 4278 that has recently become more active; however, other possibilities cannot be completely ruled out yet. The derived parameters for its jet using a one-zone synchrotron+SSC model are also significantly different from that of these typical TeV BL Lacs. In both radio and TeV bands, NGC 4278 shows luminosity lower by over two orders of magnitude when compared to these typical TeV BL Lacs. In the $R_{\rm Edd}-L_{\rm R}$ plane, NGC 4287 slightly deviates from the low-luminosity and low-Eddington-ratio extension of the 95\% confidence band representing best linear fit for BL Lacs; instead it exhibits similarities with Seyfert galaxies and LINERs.

\acknowledgments
We thank the anonymous referee for the valuable suggestions and comments. We acknowledge the use in our research the Radio Fundamental Catalogue available at \url{https://astrogeo.smce.nasa.gov/rfc/}. This work is supported by the National Key R\&D Program of China (grant 2023YFE0117200) and the National Natural Science Foundation of China (grants 12203022, 12022305, and 11973050).

\clearpage

\bibliography{biblist}{}

\begin{thebibliography}{}
\expandafter\ifx\csname natexlab\endcsname\relax\def\natexlab#1{#1}\fi
\providecommand{\url}[1]{\href{#1}{#1}}
\providecommand{\dodoi}[1]{doi:~\href{http://doi.org/#1}{\nolinkurl{#1}}}
\providecommand{\doeprint}[1]{\href{http://ascl.net/#1}{\nolinkurl{http://ascl.net/#1}}}
\providecommand{\doarXiv}[1]{\href{https://arxiv.org/abs/#1}{\nolinkurl{https://arxiv.org/abs/#1}}}

\bibitem[{{Abdo} {et~al.}(2009){Abdo}, {Ackermann}, {Ajello}, {Baldini},
  {Ballet}, {Barbiellini}, {Bastieri}, {Bechtol}, {Bellazzini}, {Berenji},
  {Bloom}, {Bonamente}, {Borgland}, {Bregeon}, {Brez}, {Brigida}, {Bruel},
  {Burnett}, {Caliandro}, {Cameron}, {Caraveo}, {Casandjian}, {Cecchi},
  {{\c{C}}elik}, {Chekhtman}, {Cheung}, {Chiang}, {Ciprini}, {Claus},
  {Cohen-Tanugi}, {Conrad}, {Cutini}, {Dermer}, {de Palma}, {Silva}, {Drell},
  {Dubois}, {Dumora}, {Farnier}, {Favuzzi}, {Fegan}, {Focke}, {Foschini},
  {Frailis}, {Fukazawa}, {Fusco}, {Gargano}, {Gehrels}, {Germani}, {Giebels},
  {Giglietto}, {Giordano}, {Giroletti}, {Glanzman}, {Godfrey}, {Grenier},
  {Grove}, {Guillemot}, {Guiriec}, {Hayashida}, {Hays}, {Horan}, {Hughes},
  {J{\'o}hannesson}, {Johnson}, {Johnson}, {Kadler}, {Kamae}, {Katagiri},
  {Kataoka}, {Kerr}, {Kn{\"o}dlseder}, {Kuss}, {Lande}, {Latronico}, {Longo},
  {Loparco}, {Lott}, {Lovellette}, {Lubrano}, {Makeev}, {Mazziotta},
  {McConville}, {McEnery}, {Meurer}, {Michelson}, {Mitthumsiri}, {Mizuno},
  {Monte}, {Monzani}, {Morselli}, {Moskalenko}, {Murgia}, {Nolan}, {Norris},
  {Nuss}, {Ohsugi}, {Omodei}, {Orlando}, {Ormes}, {Pelassa}, {Pepe}, {Persic},
  {Pesce-Rollins}, {Piron}, {Porter}, {Rain{\`o}}, {Rando}, {Razzano},
  {Rochester}, {Rodriguez}, {Ryde}, {Sadrozinski}, {Sambruna}, {Sander}, {Saz
  Parkinson}, {Scargle}, {Sgr{\`o}}, {Smith}, {Spandre}, {Spinelli},
  {Strickman}, {Suson}, {Tagliaferri}, {Takahashi}, {Takahashi}, {Tanaka},
  {Thayer}, {Thayer}, {Thompson}, {Tibaldo}, {Tibolla}, {Torres}, {Tosti},
  {Tramacere}, {Uchiyama}, {Usher}, {Vasileiou}, {Vilchez}, {Vitale}, {Waite},
  {Wang}, {Winer}, {Wood}, {Ylinen}, {Ziegler}, {Fermi/LAT Collaboration},
  {Ghisellini}, {Maraschi}, \& {Tavecchio}}]{2009ApJ...707L.142A}
{Abdo}, A.~A., {Ackermann}, M., {Ajello}, M., {et~al.} 2009, \apjl, 707, L142,
  \dodoi{10.1088/0004-637X/707/2/L142}

\bibitem[{{Abdo} {et~al.}(2010{\natexlab{a}}){Abdo}, {Ackermann}, {Ajello},
  {Atwood}, {Baldini}, {Ballet}, {Barbiellini}, {Bastieri}, {Baughman},
  {Bechtol}, {Bellazzini}, {Berenji}, {Blandford}, {Bloom}, {Bonamente},
  {Borgland}, {Bregeon}, {Brez}, {Brigida}, {Bruel}, {Burnett}, {Buson},
  {Caliandro}, {Cameron}, {Caraveo}, {Casandjian}, {Cavazzuti}, {Cecchi},
  {{\c{C}}elik}, {Chekhtman}, {Cheung}, {Chiang}, {Ciprini}, {Claus},
  {Cohen-Tanugi}, {Colafrancesco}, {Cominsky}, {Conrad}, {Costamante},
  {Cutini}, {Davis}, {Dermer}, {de Angelis}, {de Palma}, {Digel}, {do Couto e
  Silva}, {Drell}, {Dubois}, {Dumora}, {Farnier}, {Favuzzi}, {Fegan}, {Finke},
  {Focke}, {Fortin}, {Fukazawa}, {Funk}, {Fusco}, {Gargano}, {Gasparrini},
  {Gehrels}, {Georganopoulos}, {Germani}, {Giebels}, {Giglietto}, {Giordano},
  {Giroletti}, {Glanzman}, {Godfrey}, {Grenier}, {Grove}, {Guillemot},
  {Guiriec}, {Hanabata}, {Harding}, {Hayashida}, {Hays}, {Hughes}, {Jackson},
  {J{\'o}hannesson G.}, {Johnson}, {Johnson}, {Johnson}, {Kamae}, {Katagiri},
  {Kataoka}, {Kawai}, {Kerr}, {Kn{\"o}dlseder}, {Kocian}, {Kuss}, {Lande},
  {Latronico}, {Lemoine-Goumard}, {Longo}, {Loparco}, {Lott}, {Lovellette},
  {Lubrano}, {Madejski}, {Makeev}, {Mazziotta}, {McConville}, {McEnery},
  {Meurer}, {Michelson}, {Mitthumsiri}, {Mizuno}, {Moiseev}, {Monte},
  {Monzani}, {Morselli}, {Moskalenko}, {Murgia}, {Nolan}, {Norris}, {Nuss},
  {Ohsugi}, {Omodei}, {Orlando}, {Ormes}, {Paneque}, {Parent}, {Pelassa},
  {Pepe}, {Pesce-Rollins}, {Piron}, {Porter}, {Rain{\`o}}, {Rando}, {Razzano},
  {Razzaque}, {Reimer}, {Reimer}, {Reposeur}, {Ritz}, {Rochester}, {Rodriguez},
  {Romani}, {Roth}, {Ryde}, {Sadrozinski}, {Sambruna}, {Sanchez}, {Sander},
  {Saz Parkinson}, {Scargle}, {Sgr{\`o}}, {Siskind}, {Smith}, {Smith},
  {Spandre}, {Spinelli}, {Starck}, {Stawarz}, {Strickman}, {Suson}, {Tajima},
  {Takahashi}, {Takahashi}, {Tanaka}, {Thayer}, {Thayer}, {Thompson},
  {Tibaldo}, {Torres}, {Tosti}, {Tramacere}, {Uchiyama}, {Vasileiou},
  {Vilchez}, {Vitale}, {Waite}, {Wallace}, {Wang}, {Winer}, {Wood}, {Ylinen},
  {Ziegler}, {Hardcastle}, {Kazanas}, \& {Fermi LAT
  Collaboration}}]{2010Sci...328..725A}
---. 2010{\natexlab{a}}, Science, 328, 725, \dodoi{10.1126/science.1184656}

\bibitem[{{Abdo} {et~al.}(2010{\natexlab{b}}){Abdo}, {Ackermann}, {Ajello},
  {Allafort}, {Antolini}, {Atwood}, {Axelsson}, {Baldini}, {Ballet},
  {Barbiellini}, {Bastieri}, {Baughman}, {Bechtol}, {Bellazzini}, {Belli},
  {Berenji}, {Bisello}, {Blandford}, {Bloom}, {Bonamente}, {Bonnell},
  {Borgland}, {Bouvier}, {Bregeon}, {Brez}, {Brigida}, {Bruel}, {Burnett},
  {Busetto}, {Buson}, {Caliandro}, {Cameron}, {Campana}, {Canadas}, {Caraveo},
  {Carrigan}, {Casandjian}, {Cavazzuti}, {Ceccanti}, {Cecchi}, {{\c{C}}elik},
  {Charles}, {Chekhtman}, {Cheung}, {Chiang}, {Cillis}, {Ciprini}, {Claus},
  {Cohen-Tanugi}, {Conrad}, {Corbet}, {Davis}, {DeKlotz}, {den Hartog},
  {Dermer}, {de Angelis}, {de Luca}, {de Palma}, {Digel}, {Dormody}, {Silva},
  {Drell}, {Dubois}, {Dumora}, {Fabiani}, {Farnier}, {Favuzzi}, {Fegan},
  {Ferrara}, {Focke}, {Fortin}, {Frailis}, {Fukazawa}, {Funk}, {Fusco},
  {Gargano}, {Gasparrini}, {Gehrels}, {Germani}, {Giavitto}, {Giebels},
  {Giglietto}, {Giommi}, {Giordano}, {Giroletti}, {Glanzman}, {Godfrey},
  {Grenier}, {Grondin}, {Grove}, {Guillemot}, {Guiriec}, {Gustafsson},
  {Hadasch}, {Hanabata}, {Harding}, {Hayashida}, {Hays}, {Healey}, {Hill},
  {Horan}, {Hughes}, {Iafrate}, {J{\'o}hannesson}, {Johnson}, {Johnson},
  {Johnson}, {Johnson}, {Kamae}, {Katagiri}, {Kataoka}, {Kawai}, {Kerr},
  {Kn{\"o}dlseder}, {Kocevski}, {Kuss}, {Lande}, {Landriu}, {Latronico}, {Lee},
  {Lemoine-Goumard}, {Lionetto}, {Llena Garde}, {Longo}, {Loparco}, {Lott},
  {Lovellette}, {Lubrano}, {Madejski}, {Makeev}, {Marangelli}, {Marelli},
  {Massaro}, {Mazziotta}, {McConville}, {McEnery}, {Michelson}, {Minuti},
  {Mitthumsiri}, {Mizuno}, {Moiseev}, {Mongelli}, {Monte}, {Monzani},
  {Moretti}, {Morselli}, {Moskalenko}, {Murgia}, {Nakajima}, {Nakamori},
  {Naumann-Godo}, {Nolan}, {Norris}, {Nuss}, {Ohno}, {Ohsugi}, {Omodei},
  {Orlando}, {Ormes}, {Ozaki}, {Paccagnella}, {Paneque}, {Panetta}, {Parent},
  {Pelassa}, {Pepe}, {Pesce-Rollins}, {Pinchera}, {Piron}, {Porter}, {Poupard},
  {Rain{\`o}}, {Rando}, {Ray}, {Razzano}, {Razzaque}, {Rea}, {Reimer},
  {Reimer}, {Reposeur}, {Ripken}, {Ritz}, {Rochester}, {Rodriguez}, {Romani},
  {Roth}, {Sadrozinski}, {Salvetti}, {Sanchez}, {Sander}, {Saz Parkinson},
  {Scargle}, {Schalk}, {Scolieri}, {Sgr{\`o}}, {Shaw}, {Siskind}, {Smith},
  {Smith}, {Spandre}, {Spinelli}, {Starck}, {Stephens}, {Striani}, {Strickman},
  {Strong}, {Suson}, {Tajima}, {Takahashi}, {Takahashi}, {Tanaka}, {Thayer},
  {Thayer}, {Thompson}, {Tibaldo}, {Tibolla}, {Tinebra}, {Torres}, {Tosti},
  {Tramacere}, {Uchiyama}, {Usher}, {Van Etten}, {Vasileiou}, {Vilchez},
  {Vitale}, {Waite}, {Wallace}, {Wang}, {Watters}, {Winer}, {Wood}, {Yang},
  {Ylinen}, {Ziegler}, \& {Fermi LAT Collaboration}}]{2010ApJS..188..405A}
---. 2010{\natexlab{b}}, \apjs, 188, 405, \dodoi{10.1088/0067-0049/188/2/405}

\bibitem[{{Abdollahi} {et~al.}(2020){Abdollahi}, {Acero}, {Ackermann},
  {Ajello}, {Atwood}, {Axelsson}, {Baldini}, {Ballet}, {Barbiellini},
  {Bastieri}, {Becerra Gonzalez}, {Bellazzini}, {Berretta}, {Bissaldi},
  {Blandford}, {Bloom}, {Bonino}, {Bottacini}, {Brandt}, {Bregeon}, {Bruel},
  {Buehler}, {Burnett}, {Buson}, {Cameron}, {Caputo}, {Caraveo}, {Casandjian},
  {Castro}, {Cavazzuti}, {Charles}, {Chaty}, {Chen}, {Cheung}, {Chiaro},
  {Ciprini}, {Cohen-Tanugi}, {Cominsky}, {Coronado-Bl{\'a}zquez}, {Costantin},
  {Cuoco}, {Cutini}, {D'Ammando}, {DeKlotz}, {de la Torre Luque}, {de Palma},
  {Desai}, {Digel}, {Di Lalla}, {Di Mauro}, {Di Venere}, {Dom{\'\i}nguez},
  {Dumora}, {Fana Dirirsa}, {Fegan}, {Ferrara}, {Franckowiak}, {Fukazawa},
  {Funk}, {Fusco}, {Gargano}, {Gasparrini}, {Giglietto}, {Giommi}, {Giordano},
  {Giroletti}, {Glanzman}, {Green}, {Grenier}, {Griffin}, {Grondin}, {Grove},
  {Guiriec}, {Harding}, {Hayashi}, {Hays}, {Hewitt}, {Horan},
  {J{\'o}hannesson}, {Johnson}, {Kamae}, {Kerr}, {Kocevski}, {Kovac'evic'},
  {Kuss}, {Landriu}, {Larsson}, {Latronico}, {Lemoine-Goumard}, {Li},
  {Liodakis}, {Longo}, {Loparco}, {Lott}, {Lovellette}, {Lubrano}, {Madejski},
  {Maldera}, {Malyshev}, {Manfreda}, {Marchesini}, {Marcotulli},
  {Mart{\'\i}-Devesa}, {Martin}, {Massaro}, {Mazziotta}, {McEnery}, {Mereu},
  {Meyer}, {Michelson}, {Mirabal}, {Mizuno}, {Monzani}, {Morselli},
  {Moskalenko}, {Negro}, {Nuss}, {Ojha}, {Omodei}, {Orienti}, {Orlando},
  {Ormes}, {Palatiello}, {Paliya}, {Paneque}, {Pei}, {Pe{\~n}a-Herazo},
  {Perkins}, {Persic}, {Pesce-Rollins}, {Petrosian}, {Petrov}, {Piron}, {Poon},
  {Porter}, {Principe}, {Rain{\`o}}, {Rando}, {Razzano}, {Razzaque}, {Reimer},
  {Reimer}, {Remy}, {Reposeur}, {Romani}, {Saz Parkinson}, {Schinzel},
  {Serini}, {Sgr{\`o}}, {Siskind}, {Smith}, {Spandre}, {Spinelli}, {Strong},
  {Suson}, {Tajima}, {Takahashi}, {Tak}, {Thayer}, {Thompson}, {Tibaldo},
  {Torres}, {Torresi}, {Valverde}, {Van Klaveren}, {van Zyl}, {Wood},
  {Yassine}, \& {Zaharijas}}]{2020ApJS..247...33A}
{Abdollahi}, S., {Acero}, F., {Ackermann}, M., {et~al.} 2020, \apjs, 247, 33,
  \dodoi{10.3847/1538-4365/ab6bcb}

\bibitem[{{Abdollahi} {et~al.}(2022){Abdollahi}, {Acero}, {Baldini}, {Ballet},
  {Bastieri}, {Bellazzini}, {Berenji}, {Berretta}, {Bissaldi}, {Blandford},
  {Bloom}, {Bonino}, {Brill}, {Britto}, {Bruel}, {Burnett}, {Buson}, {Cameron},
  {Caputo}, {Caraveo}, {Castro}, {Chaty}, {Cheung}, {Chiaro}, {Cibrario},
  {Ciprini}, {Coronado-Bl{\'a}zquez}, {Crnogorcevic}, {Cutini}, {D'Ammando},
  {De Gaetano}, {Digel}, {Di Lalla}, {Dirirsa}, {Di Venere}, {Dom{\'\i}nguez},
  {Fallah Ramazani}, {Fegan}, {Ferrara}, {Fiori}, {Fleischhack}, {Franckowiak},
  {Fukazawa}, {Funk}, {Fusco}, {Galanti}, {Gammaldi}, {Gargano}, {Garrappa},
  {Gasparrini}, {Giacchino}, {Giglietto}, {Giordano}, {Giroletti}, {Glanzman},
  {Green}, {Grenier}, {Grondin}, {Guillemot}, {Guiriec}, {Gustafsson},
  {Harding}, {Hays}, {Hewitt}, {Horan}, {Hou}, {J{\'o}hannesson}, {Karwin},
  {Kayanoki}, {Kerr}, {Kuss}, {Landriu}, {Larsson}, {Latronico},
  {Lemoine-Goumard}, {Li}, {Liodakis}, {Longo}, {Loparco}, {Lott}, {Lubrano},
  {Maldera}, {Malyshev}, {Manfreda}, {Mart{\'\i}-Devesa}, {Mazziotta}, {Mereu},
  {Meyer}, {Michelson}, {Mirabal}, {Mitthumsiri}, {Mizuno}, {Moiseev},
  {Monzani}, {Morselli}, {Moskalenko}, {Negro}, {Nuss}, {Omodei}, {Orienti},
  {Orlando}, {Paneque}, {Pei}, {Perkins}, {Persic}, {Pesce-Rollins},
  {Petrosian}, {Pillera}, {Poon}, {Porter}, {Principe}, {Rain{\`o}}, {Rando},
  {Rani}, {Razzano}, {Razzaque}, {Reimer}, {Reimer}, {Reposeur},
  {S{\'a}nchez-Conde}, {Saz Parkinson}, {Scotton}, {Serini}, {Sgr{\`o}},
  {Siskind}, {Smith}, {Spandre}, {Spinelli}, {Sueoka}, {Suson}, {Tajima},
  {Tak}, {Thayer}, {Thompson}, {Torres}, {Troja}, {Valverde}, {Wood}, \&
  {Zaharijas}}]{2022ApJS..260...53A}
{Abdollahi}, S., {Acero}, F., {Baldini}, L., {et~al.} 2022, \apjs, 260, 53,
  \dodoi{10.3847/1538-4365/ac6751}

\bibitem[{{Abeysekara} {et~al.}(2017){Abeysekara}, {Archambault}, {Archer},
  {Benbow}, {Bird}, {Buchovecky}, {Buckley}, {Bugaev}, {Byrum}, {Cerruti},
  {Chen}, {Ciupik}, {Cui}, {Dickinson}, {Eisch}, {Errando}, {Falcone}, {Feng},
  {Finley}, {Fleischhack}, {Fortson}, {Furniss}, {Gillanders}, {Griffin},
  {Grube}, {H{\"u}tten}, {H{\r{a}}kansson}, {Hanna}, {Holder}, {Humensky},
  {Johnson}, {Kaaret}, {Kar}, {Kertzman}, {Kieda}, {Krause}, {Krennrich},
  {Kumar}, {Lang}, {Maier}, {McArthur}, {McCann}, {Meagher}, {Moriarty},
  {Mukherjee}, {Nguyen}, {Nieto}, {Ong}, {Otte}, {Park}, {Pelassa}, {Pohl},
  {Popkow}, {Pueschel}, {Quinn}, {Ragan}, {Reynolds}, {Richards}, {Roache},
  {Rulten}, {Santander}, {Sembroski}, {Shahinyan}, {Staszak}, {Telezhinsky},
  {Tucci}, {Tyler}, {Wakely}, {Weiner}, {Weinstein}, {Wilhelm}, {Williams},
  {VERITAS Collaboration}, {Fegan}, {Giebels}, {Horan}, {Fermi-LAT
  Collaboration}, {Berdyugin}, {Kuan}, {Lindfors}, {Nilsson}, {Oksanen},
  {Prokoph}, {Reinthal}, {Takalo}, \& {Zefi}}]{2017ApJ...836..205A}
{Abeysekara}, A.~U., {Archambault}, S., {Archer}, A., {et~al.} 2017, \apj, 836,
  205, \dodoi{10.3847/1538-4357/836/2/205}

\bibitem[{{Acciari} {et~al.}(2009){Acciari}, {Aliu}, {Arlen}, {Beilicke},
  {Benbow}, {Bradbury}, {Buckley}, {Bugaev}, {Butt}, {Byrum}, {Celik},
  {Cesarini}, {Ciupik}, {Chow}, {Cogan}, {Colin}, {Cui}, {Daniel}, {Ergin},
  {Falcone}, {Fegan}, {Finley}, {Fortin}, {Fortson}, {Furniss}, {Gillanders},
  {Grube}, {Guenette}, {Gyuk}, {Hanna}, {Hays}, {Holder}, {Horan}, {Hui},
  {Humensky}, {Imran}, {Kaaret}, {Karlsson}, {Kertzman}, {Kieda}, {Kildea},
  {Konopelko}, {Krawczynski}, {Krennrich}, {Lang}, {LeBohec}, {Maier},
  {McCann}, {McCutcheon}, {Moriarty}, {Mukherjee}, {Nagai}, {Niemiec}, {Ong},
  {Pandel}, {Perkins}, {Pohl}, {Quinn}, {Ragan}, {Reyes}, {Reynolds}, {Rose},
  {Schroedter}, {Sembroski}, {Smith}, {Steele}, {Swordy}, {Toner}, {Valcarcel},
  {Vassiliev}, {Wagner}, {Wakely}, {Ward}, {Weekes}, {Weinstein}, {White},
  {Williams}, {Wissel}, {Wood}, \& {Zitzer}}]{2009ApJ...695.1370A}
{Acciari}, V.~A., {Aliu}, E., {Arlen}, T., {et~al.} 2009, \apj, 695, 1370,
  \dodoi{10.1088/0004-637X/695/2/1370}

\bibitem[{{Ackermann} {et~al.}(2016){Ackermann}, {Ajello}, {Baldini}, {Ballet},
  {Barbiellini}, {Bastieri}, {Bellazzini}, {Bissaldi}, {Blandford}, {Bloom},
  {Bonino}, {Brandt}, {Bregeon}, {Bruel}, {Buehler}, {Buson}, {Caliandro},
  {Cameron}, {Caragiulo}, {Caraveo}, {Cavazzuti}, {Cecchi}, {Charles},
  {Chekhtman}, {Cheung}, {Chiaro}, {Ciprini}, {Cohen}, {Cohen-Tanugi},
  {Costanza}, {Cutini}, {D'Ammando}, {Davis}, {de Angelis}, {de Palma},
  {Desiante}, {Digel}, {Di Lalla}, {Di Mauro}, {Di Venere}, {Favuzzi}, {Fegan},
  {Ferrara}, {Focke}, {Fukazawa}, {Funk}, {Fusco}, {Gargano}, {Gasparrini},
  {Georganopoulos}, {Giglietto}, {Giordano}, {Giroletti}, {Godfrey}, {Green},
  {Grenier}, {Guiriec}, {Hays}, {Hewitt}, {Hill}, {Jogler}, {J{\'o}hannesson},
  {Kensei}, {Kuss}, {Larsson}, {Latronico}, {Li}, {Li}, {Longo}, {Loparco},
  {Lubrano}, {Magill}, {Maldera}, {Manfreda}, {Mayer}, {Mazziotta},
  {McConville}, {McEnery}, {Michelson}, {Mitthumsiri}, {Mizuno}, {Monzani},
  {Morselli}, {Moskalenko}, {Murgia}, {Negro}, {Nuss}, {Ohno}, {Ohsugi},
  {Orienti}, {Orlando}, {Ormes}, {Paneque}, {Perkins}, {Pesce-Rollins},
  {Piron}, {Pivato}, {Porter}, {Rain{\`o}}, {Rando}, {Razzano}, {Reimer},
  {Reimer}, {Schmid}, {Sgr{\`o}}, {Simone}, {Siskind}, {Spada}, {Spandre},
  {Spinelli}, {Stawarz}, {Takahashi}, {Thayer}, {Thompson}, {Torres}, {Tosti},
  {Troja}, {Vianello}, {Wood}, {Wood}, {Zimmer}, \& {Fermi LAT
  Collaboration}}]{2016ApJ...826....1A}
{Ackermann}, M., {Ajello}, M., {Baldini}, L., {et~al.} 2016, \apj, 826, 1,
  \dodoi{10.3847/0004-637X/826/1/1}

\bibitem[{{Akiyama} {et~al.}(2003){Akiyama}, {Ueda}, {Ohta}, {Takahashi}, \&
  {Yamada}}]{2003ApJS..148..275A}
{Akiyama}, M., {Ueda}, Y., {Ohta}, K., {Takahashi}, T., \& {Yamada}, T. 2003,
  \apjs, 148, 275, \dodoi{10.1086/376441}

\bibitem[{{Albert} {et~al.}(2006){Albert}, {Aliu}, {Anderhub}, {Antoranz},
  {Armada}, {Asensio}, {Baixeras}, {Barrio}, {Bartelt}, {Bartko}, {Bastieri},
  {Bavikadi}, {Bednarek}, {Berger}, {Bigongiari}, {Biland}, {Bisesi}, {Bock},
  {Bretz}, {Britvitch}, {Camara}, {Chilingarian}, {Ciprini}, {Coarasa},
  {Commichau}, {Contreras}, {Cortina}, {Curtef}, {Danielyan}, {Dazzi}, {De
  Angelis}, {de los Reyes}, {De Lotto}, {Domingo-Santamar{\'\i}a}, {Dorner},
  {Doro}, {Errando}, {Fagiolini}, {Ferenc}, {Fern{\'a}ndez}, {Firpo}, {Flix},
  {Fonseca}, {Font}, {Galante}, {Garczarczyk}, {Gaug}, {Giller}, {Goebel},
  {Hakobyan}, {Hayashida}, {Hengstebeck}, {H{\"o}hne}, {Hose}, {Jacon},
  {Kalekin}, {Kranich}, {Laille}, {Lenisa}, {Liebing}, {Lindfors}, {Longo},
  {L{\'o}pez}, {L{\'o}pez}, {Lorenz}, {Lucarelli}, {Majumdar}, {Maneva},
  {Mannheim}, {Mariotti}, {Mart{\'\i}nez}, {Mase}, {Mazin}, {Meucci}, {Meyer},
  {Miranda}, {Mirzoyan}, {Mizobuchi}, {Moralejo}, {Nilsson},
  {O{\~n}a-Wilhelmi}, {Ordu{\~n}a}, {Otte}, {Oya}, {Paneque}, {Paoletti},
  {Pasanen}, {Pascoli}, {Pauss}, {Pavel}, {Pegna}, {Persic}, {Peruzzo},
  {Piccioli}, {Poller}, {Prandini}, {Rhode}, {Rico}, {Riegel}, {Rissi},
  {Robert}, {R{\"u}gamer}, {Saggion}, {S{\'a}nchez}, {Sartori}, {Scalzotto},
  {Schmitt}, {Schweizer}, {Shayduk}, {Shinozaki}, {Shore}, {Sidro},
  {Sillanp{\"a}{\"a}}, {Sobczy{\'n}ska}, {Stamerra}, {Stark}, {Takalo},
  {Temnikov}, {Tescaro}, {Teshima}, {Tonello}, {Torres}, {Torres}, {Turini},
  {Vankov}, {Vardanyan}, {Vitale}, {Wagner}, {Wibig}, {Wittek}, \&
  {Zapatero}}]{2006ApJ...642L.119A}
{Albert}, J., {Aliu}, E., {Anderhub}, H., {et~al.} 2006, \apjl, 642, L119,
  \dodoi{10.1086/504845}

\bibitem[{{Aleksi{\'c}} {et~al.}(2012{\natexlab{a}}){Aleksi{\'c}}, {Alvarez},
  {Antonelli}, {Antoranz}, {Ansoldi}, {Asensio}, {Backes}, {Barres de Almeida},
  {Barrio}, {Bastieri}, {Becerra Gonz{\'a}lez}, {Bednarek}, {Berger},
  {Bernardini}, {Biland}, {Blanch}, {Bock}, {Boller}, {Bonnoli}, {Borla
  Tridon}, {Bretz}, {Ca{\~n}ellas}, {Carmona}, {Carosi}, {Colin}, {Colombo},
  {Contreras}, {Cortina}, {Cossio}, {Covino}, {Da Vela}, {Dazzi}, {De Angelis},
  {De Caneva}, {De Cea del Pozo}, {De Lotto}, {Delgado Mendez}, {Diago Ortega},
  {Doert}, {Dom{\'\i}nguez}, {Dominis Prester}, {Dorner}, {Doro}, {Eisenacher},
  {Elsaesser}, {Ferenc}, {Fonseca}, {Font}, {Fruck}, {Garc{\'\i}a L{\'o}pez},
  {Garczarczyk}, {Garrido Terrats}, {Gaug}, {Giavitto}, {Godinovi{\'c}},
  {Gonz{\'a}lez Mu{\~n}oz}, {Gozzini}, {Hadasch}, {H{\"a}fner}, {Herrero},
  {Hildebrand}, {Hose}, {Hrupec}, {Huber}, {Jankowski}, {Jogler}, {Kadenius},
  {Kellermann}, {Klepser}, {Kr{\"a}henb{\"u}hl}, {Krause}, {La Barbera},
  {Lelas}, {Leonardo}, {Lewandowska}, {Lindfors}, {Lombardi}, {L{\'o}pez},
  {L{\'o}pez-Coto}, {L{\'o}pez-Oramas}, {Lorenz}, {Makariev}, {Maneva},
  {Mankuzhiyil}, {Mannheim}, {Maraschi}, {Mariotti}, {Mart{\'\i}nez}, {Mazin},
  {Meucci}, {Miranda}, {Mirzoyan}, {Mold{\'o}n}, {Moralejo}, {Munar-Adrover},
  {Niedzwiecki}, {Nieto}, {Nilsson}, {Nowak}, {Orito}, {Paiano}, {Paneque},
  {Paoletti}, {Pardo}, {Paredes}, {Partini}, {Perez-Torres}, {Persic}, {Pilia},
  {Pochon}, {Prada}, {Prada Moroni}, {Prandini}, {Puerto Gimenez}, {Puljak},
  {Reichardt}, {Reinthal}, {Rhode}, {Rib{\'o}}, {Rico}, {R{\"u}gamer},
  {Saggion}, {Saito}, {Saito}, {Salvati}, {Satalecka}, {Scalzotto}, {Scapin},
  {Schultz}, {Schweizer}, {Shore}, {Sillanp{\"a}{\"a}}, {Sitarek}, {Snidaric},
  {Sobczynska}, {Spanier}, {Spiro}, {Stamatescu}, {Stamerra}, {Steinke},
  {Storz}, {Strah}, {Sun}, {Suri{\'c}}, {Takalo}, {Takami}, {Tavecchio},
  {Temnikov}, {Terzi{\'c}}, {Tescaro}, {Teshima}, {Tibolla}, {Torres},
  {Treves}, {Uellenbeck}, {Vogler}, {Wagner}, {Weitzel}, {Zabalza}, {Zandanel},
  {Zanin}, {Berdyugin}, {Buson}, {J{\"a}rvel{\"a}}, {Larsson},
  {L{\"a}hteenm{\"a}ki}, \& {Tammi}}]{2012A&A...544A.142A}
{Aleksi{\'c}}, J., {Alvarez}, E.~A., {Antonelli}, L.~A., {et~al.}
  2012{\natexlab{a}}, \aap, 544, A142, \dodoi{10.1051/0004-6361/201219133}

\bibitem[{{Aleksi{\'c}} {et~al.}(2012{\natexlab{b}}){Aleksi{\'c}}, {Alvarez},
  {Antonelli}, {Antoranz}, {Asensio}, {Backes}, {Barres de Almeida}, {Barrio},
  {Bastieri}, {Becerra Gonz{\'a}lez}, {Bednarek}, {Berger}, {Bernardini},
  {Biland}, {Blanch}, {Bock}, {Boller}, {Bonnoli}, {Borla Tridon}, {Bretz},
  {Ca{\~n}ellas}, {Carmona}, {Carosi}, {Colin}, {Colombo}, {Contreras},
  {Cortina}, {Cossio}, {Covino}, {da Vela}, {Dazzi}, {de Angelis}, {de Caneva},
  {de Cea Del Pozo}, {de Lotto}, {Delgado Mendez}, {Diago Ortega}, {Doert},
  {Dom{\'\i}nguez}, {Dominis Prester}, {Dorner}, {Doro}, {Eisenacher},
  {Elsaesser}, {Ferenc}, {Fonseca}, {Font}, {Fruck}, {Garc{\'\i}a L{\'o}pez},
  {Garczarczyk}, {Garrido}, {Giavitto}, {Godinovi{\'c}}, {Gozzini}, {Hadasch},
  {H{\"a}fner}, {Herrero}, {Hildebrand}, {H{\"o}hne-M{\"o}nch}, {Hose},
  {Hrupec}, {Huber}, {Jogler}, {Kadenius}, {Kellermann}, {Klepser},
  {Kr{\"a}henb{\"u}hl}, {Krause}, {La Barbera}, {Lelas}, {Leonardo},
  {Lewandowska}, {Lindfors}, {Lombardi}, {L{\'o}pez}, {L{\'o}pez-Coto},
  {L{\'o}pez-Oramas}, {Lorenz}, {Makariev}, {Maneva}, {Mankuzhiyil},
  {Mannheim}, {Maraschi}, {Mariotti}, {Mart{\'\i}nez}, {Mazin}, {Meucci},
  {Miranda}, {Mirzoyan}, {Mold{\'o}n}, {Moralejo}, {Munar-Adrover},
  {Niedzwiecki}, {Nieto}, {Nilsson}, {Nowak}, {Orito}, {Paiano}, {Paneque},
  {Paoletti}, {Pardo}, {Paredes}, {Partini}, {Perez-Torres}, {Persic},
  {Peruzzo}, {Pilia}, {Pochon}, {Prada}, {Prada Moroni}, {Prandini}, {Puerto
  Gimenez}, {Puljak}, {Reichardt}, {Reinthal}, {Rhode}, {Rib{\'o}}, {Rico},
  {R{\"u}gamer}, {Saggion}, {Saito}, {Saito}, {Salvati}, {Satalecka},
  {Scalzotto}, {Scapin}, {Schultz}, {Schweizer}, {Shayduk}, {Shore},
  {Sillanp{\"a}{\"a}}, {Sitarek}, {Snidaric}, {Sobczynska}, {Spanier}, {Spiro},
  {Stamatescu}, {Stamerra}, {Steinke}, {Storz}, {Strah}, {Sun}, {Suri{\'c}},
  {Takalo}, {Takami}, {Tavecchio}, {Temnikov}, {Terzi{\'c}}, {Tescaro},
  {Teshima}, {Tibolla}, {Torres}, {Treves}, {Uellenbeck}, {Vogler}, {Wagner},
  {Weitzel}, {Zabalza}, {Zandanel}, {Zanin}, {Pfrommer}, \&
  {Pinzke}}]{2012A&A...539L...2A}
---. 2012{\natexlab{b}}, \aap, 539, L2, \dodoi{10.1051/0004-6361/201118668}

\bibitem[{{Aliu} {et~al.}(2013){Aliu}, {Archambault}, {Arlen}, {Aune},
  {Beilicke}, {Benbow}, {Bird}, {Bouvier}, {Buckley}, {Bugaev}, {Cesarini},
  {Ciupik}, {Connolly}, {Cui}, {Dumm}, {Errando}, {Falcone}, {Federici},
  {Feng}, {Finley}, {Fortin}, {Fortson}, {Furniss}, {Galante}, {G{\'e}rard},
  {Gillanders}, {Griffin}, {Grube}, {Gyuk}, {Hanna}, {Holder}, {Hughes},
  {Humensky}, {Kaaret}, {Kertzman}, {Khassen}, {Kieda}, {Krawczynski},
  {Krennrich}, {Lang}, {Madhavan}, {Maier}, {Majumdar}, {McArthur}, {McCann},
  {Moriarty}, {Mukherjee}, {Nieto}, {O'Faol{\'a}in de Bhr{\'o}ithe}, {Ong},
  {Orr}, {Otte}, {Park}, {Perkins}, {Pohl}, {Popkow}, {Prokoph}, {Quinn},
  {Ragan}, {Reyes}, {Reynolds}, {Richards}, {Roache}, {Saxon}, {Sembroski},
  {Skole}, {Smith}, {Soares-Furtado}, {Staszak}, {Telezhinsky},
  {Te{\v{s}}i{\'c}}, {Theiling}, {Varlotta}, {Vassiliev}, {Vincent}, {Wakely},
  {Weekes}, {Weinstein}, {Welsing}, {Williams}, {Zitzer}, {VERITAS
  Collaboration}, {B{\"o}ttcher}, {Fumagalli}, \&
  {Jadhav}}]{2013ApJ...779...92A}
{Aliu}, E., {Archambault}, S., {Arlen}, T., {et~al.} 2013, \apj, 779, 92,
  \dodoi{10.1088/0004-637X/779/2/92}

\bibitem[{{Baldini} {et~al.}(2021){Baldini}, {Ballet}, {Bastieri}, {Becerra
  Gonzalez}, {Bellazzini}, {Berretta}, {Bissaldi}, {Blandford}, {Bloom},
  {Bonino}, {Bottacini}, {Bruel}, {Buson}, {Cameron}, {Caraveo}, {Cavazzuti},
  {Chen}, {Chiaro}, {Ciangottini}, {Cibario}, {Ciprini}, {Cristarella
  Orestano}, {Crnogorcevic}, {Cutini}, {D'Ammando}, {de la Torre Luque}, {de
  Palma}, {Digel}, {Di Lalla}, {Dirirsa}, {Di Venere}, {Dom{\'\i}nguez},
  {Fiori}, {Fleischhack}, {Franckowiak}, {Fukazawa}, {Funk}, {Fusco},
  {Gargano}, {Gasparrini}, {Germani}, {Giglietto}, {Giordano}, {Giroletti},
  {Green}, {Grenier}, {Griffin}, {Guiriec}, {Gustafsson}, {Hewitt}, {Horan},
  {Imazawa}, {J{\'o}hannesson}, {Kerr}, {Kocevski}, {Kuss}, {Larsson},
  {Latronico}, {Li}, {Liodakis}, {Longo}, {Loparco}, {Lovellette}, {Lubrano},
  {Maldera}, {Manfreda}, {Mart{\'\i}-Devesa}, {Matake}, {Mazziotta}, {Mereu},
  {Meyer}, {Mirabal}, {Mitthumsiri}, {Mizuno}, {Monzani}, {Morselli},
  {Moskalenko}, {Nagasawa}, {Negro}, {Ojha}, {Orienti}, {Orlando},
  {Palatiello}, {Paliya}, {Paneque}, {Pei}, {Persic}, {Pesce-Rollins},
  {Petrosian}, {Poon}, {Porter}, {Principe}, {Racusin}, {Rain{\`o}}, {Rando},
  {Rani}, {Razzano}, {Razzaque}, {Reimer}, {Reimer}, {Saz Parkinson},
  {Scotton}, {Serini}, {Sgr{\`o}}, {Siskind}, {Spandre}, {Spinelli}, {Suson},
  {Tajima}, {Tak}, {Torres}, {Tosti}, {Troja}, {Wood}, {Yassine}, {Zaharijas},
  \& {Fermi-LAT Collaboration}}]{2021ApJS..256...13B}
{Baldini}, L., {Ballet}, J., {Bastieri}, D., {et~al.} 2021, \apjs, 256, 13,
  \dodoi{10.3847/1538-4365/ac072a}

\bibitem[{{Ballet} {et~al.}(2023){Ballet}, {Bruel}, {Burnett}, {Lott}, \& {The
  Fermi-LAT collaboration}}]{2023arXiv230712546B}
{Ballet}, J., {Bruel}, P., {Burnett}, T.~H., {Lott}, B., \& {The Fermi-LAT
  collaboration}. 2023, arXiv e-prints, arXiv:2307.12546,
  \dodoi{10.48550/arXiv.2307.12546}

\bibitem[{{Brassington} {et~al.}(2009){Brassington}, {Fabbiano}, {Kim},
  {Zezas}, {Zepf}, {Kundu}, {Angelini}, {Davies}, {Gallagher}, {Kalogera},
  {Fragos}, {King}, {Pellegrini}, \& {Trinchieri}}]{2009ApJS..181..605B}
{Brassington}, N.~J., {Fabbiano}, G., {Kim}, D.~W., {et~al.} 2009, \apjs, 181,
  605, \dodoi{10.1088/0067-0049/181/2/605}

\bibitem[{{Bruel} {et~al.}(2018){Bruel}, {Burnett}, {Digel}, {Johannesson},
  {Omodei}, \& {Wood}}]{2018arXiv181011394B}
{Bruel}, P., {Burnett}, T.~H., {Digel}, S.~W., {et~al.} 2018, arXiv e-prints,
  arXiv:1810.11394, \dodoi{10.48550/arXiv.1810.11394}

\bibitem[{{Campana} {et~al.}(2016){Campana}, {Coti Zelati}, {Papitto}, {Rea},
  {Torres}, {Baglio}, \& {D'Avanzo}}]{2016A&A...594A..31C}
{Campana}, S., {Coti Zelati}, F., {Papitto}, A., {et~al.} 2016, \aap, 594, A31,
  \dodoi{10.1051/0004-6361/201629035}

\bibitem[{{Cao} {et~al.}(2024{\natexlab{a}}){Cao}, {Aharonian}, {An},
  {Axikegu}, {Bai}, {Bao}, {Bastieri}, {Bi}, {Bi}, {Cai}, {Cao}, {Cao}, {Cao},
  {Chang}, {Chang}, {Chen}, {Chen}, {Chen}, {Chen}, {Chen}, {Chen}, {Chen},
  {Chen}, {Chen}, {Chen}, {Chen}, {Chen}, {Cheng}, {Cheng}, {Cui}, {Cui},
  {Cui}, {Cui}, {Dai}, {Dai}, {Dai}, {Danzengluobu}, {Della Volpe}, {Dong},
  {Duan}, {Fan}, {Fan}, {Fang}, {Fang}, {Feng}, {Feng}, {Feng}, {Feng}, {Feng},
  {Gabici}, {Gao}, {Gao}, {Gao}, {Gao}, {Gao}, {Gao}, {Ge}, {Geng}, {Giacinti},
  {Gong}, {Gou}, {Gu}, {Guo}, {Guo}, {Guo}, {Guo}, {Han}, {He}, {He}, {He},
  {He}, {He}, {Heller}, {Hor}, {Hou}, {Hou}, {Hou}, {Hu}, {Hu}, {Hu}, {Huang},
  {Huang}, {Huang}, {Huang}, {Huang}, {Huang}, {Huang}, {Ji}, {Jia}, {Jia},
  {Jiang}, {Jiang}, {Jiang}, {Jin}, {Kang}, {Ke}, {Kuleshov}, {Kurinov}, {Li},
  {Li}, {Li}, {Li}, {Li}, {Li}, {Li}, {Li}, {Li}, {Li}, {Li}, {Li}, {Li}, {Li},
  {Li}, {Li}, {Li}, {Li}, {Li}, {Liang}, {Liang}, {Lin}, {Liu}, {Liu}, {Liu},
  {Liu}, {Liu}, {Liu}, {Liu}, {Liu}, {Liu}, {Liu}, {Liu}, {Liu}, {Liu}, {Liu},
  {Lu}, {Luo}, {Lv}, {Ma}, {Ma}, {Ma}, {Mao}, {Min}, {Mitthumsiri}, {Mu},
  {Nan}, {Neronov}, {Ou}, {Pang}, {Pattarakijwanich}, {Pei}, {Qi}, {Qi},
  {Qiao}, {Qin}, {Ruffolo}, {S{\'a}iz}, {Semikoz}, {Shao}, {Shao},
  {Shchegolev}, {Sheng}, {Shu}, {Song}, {Stenkin}, {Stepanov}, {Su}, {Sun},
  {Sun}, {Sun}, {Tam}, {Tang}, {Tang}, {Tian}, {Wang}, {Wang}, {Wang}, {Wang},
  {Wang}, {Wang}, {Wang}, {Wang}, {Wang}, {Wang}, {Wang}, {Wang}, {Wang},
  {Wang}, {Wang}, {Wang}, {Wang}, {Wang}, {Wang}, {Wang}, {Wang}, {Wei}, {Wei},
  {Wei}, {Wen}, {Wu}, {Wu}, {Wu}, {Wu}, {Wu}, {Xi}, {Xia}, {Xia}, {Xiang},
  {Xiao}, {Xiao}, {Xin}, {Xin}, {Xing}, {Xiong}, {Xu}, {Xu}, {Xu}, {Xu}, {Xue},
  {Yan}, {Yan}, {Yan}, {Yang}, {Yang}, {Yang}, {Yang}, {Yang}, {Yang}, {Yang},
  {Yang}, {Yang}, {Yao}, {Yao}, {Ye}, {Yin}, {Yin}, {You}, {You}, {Yu}, {Yuan},
  {Yue}, {Zeng}, {Zeng}, {Zeng}, {Zha}, {Zhang}, {Zhang}, {Zhang}, {Zhang},
  {Zhang}, {Zhang}, {Zhang}, {Zhang}, {Zhang}, {Zhang}, {Zhang}, {Zhang},
  {Zhang}, {Zhang}, {Zhang}, {Zhang}, {Zhang}, {Zhang}, {Zhao}, {Zhao}, {Zhao},
  {Zhao}, {Zhao}, {Zheng}, {Zhou}, {Zhou}, {Zhou}, {Zhou}, {Zhou}, {Zhou},
  {Zhou}, {Zhu}, {Zhu}, {Zhu}, {Zhu}, {Zuo}, \& {(The Lhaaso
  Collaboration)}}]{2024ApJS..271...25C}
{Cao}, Z., {Aharonian}, F., {An}, Q., {et~al.} 2024{\natexlab{a}}, \apjs, 271,
  25, \dodoi{10.3847/1538-4365/acfd29}

\bibitem[{{Cao} {et~al.}(2024{\natexlab{b}}){Cao}, {Aharonian}, {An},
  {Axikegu}, {Bai}, {Bao}, {Bastieri}, {Bi}, {Bi}, {Cai}, {Cao}, {Cao}, {Cao},
  {Chang}, {Chang}, {Chen}, {Chen}, {Chen}, {Chen}, {Chen}, {Chen}, {Chen},
  {Chen}, {Chen}, {Chen}, {Chen}, {Chen}, {Cheng}, {Cheng}, {Cui}, {Cui},
  {Cui}, {Cui}, {Dai}, {Dai}, {Dai}, {Danzengluobu}, {Dong}, {Duan}, {Fan},
  {Fan}, {Fang}, {Fang}, {Feng}, {Feng}, {Feng}, {Feng}, {Feng}, {Gabici},
  {Gao}, {Gao}, {Gao}, {Gao}, {Gao}, {Gao}, {Ge}, {Geng}, {Giacinti}, {Gong},
  {Gou}, {Gu}, {Guo}, {Guo}, {Guo}, {Guo}, {Han}, {He}, {He}, {He}, {He}, {He},
  {Hor}, {Hou}, {Hou}, {Hou}, {Hu}, {Hu}, {Hu}, {Huang}, {Huang}, {Huang},
  {Huang}, {Huang}, {Huang}, {Huang}, {Ji}, {Jia}, {Jia}, {Jiang}, {Jiang},
  {Jiang}, {Jin}, {Kang}, {Ke}, {Kuleshov}, {Kurinov}, {Li}, {Li}, {Li}, {Li},
  {Li}, {Li}, {Li}, {Li}, {Li}, {Li}, {Li}, {Li}, {Li}, {Li}, {Li}, {Li}, {Li},
  {Li}, {Li}, {Liang}, {Liang}, {Lin}, {Liu}, {Liu}, {Liu}, {Liu}, {Liu},
  {Liu}, {Liu}, {Liu}, {Liu}, {Liu}, {Liu}, {Liu}, {Liu}, {Liu}, {Lu}, {Luo},
  {Lv}, {Ma}, {Ma}, {Ma}, {Mao}, {Min}, {Mitthumsiri}, {Mu}, {Nan}, {Neronov},
  {Ou}, {Pang}, {Pattarakijwanich}, {Pei}, {Qi}, {Qi}, {Qiao}, {Qin},
  {Ruffolo}, {S{\'a}iz}, {Semikoz}, {Shao}, {Shao}, {Shchegolev}, {Sheng},
  {Shu}, {Song}, {Stenkin}, {Stepanov}, {Su}, {Sun}, {Sun}, {Sun}, {Tam},
  {Tang}, {Tang}, {Tian}, {Wang}, {Wang}, {Wang}, {Wang}, {Wang}, {Wang},
  {Wang}, {Wang}, {Wang}, {Wang}, {Wang}, {Wang}, {Wang}, {Wang}, {Wang},
  {Wang}, {Wang}, {Wang}, {Wang}, {Wang}, {Wang}, {Wei}, {Wei}, {Wei}, {Wen},
  {Wu}, {Wu}, {Wu}, {Wu}, {Wu}, {Xi}, {Xia}, {Xia}, {Xiang}, {Xiao}, {Xiao},
  {Xin}, {Xin}, {Xing}, {Xiong}, {Xu}, {Xu}, {Xu}, {Xu}, {Xue}, {Yan}, {Yan},
  {Yan}, {Yang}, {Yang}, {Yang}, {Yang}, {Yang}, {Yang}, {Yang}, {Yang},
  {Yang}, {Yao}, {Yao}, {Ye}, {Yin}, {Yin}, {You}, {You}, {Yu}, {Yuan}, {Yue},
  {Zeng}, {Zeng}, {Zeng}, {Zha}, {Zhang}, {Zhang}, {Zhang}, {Zhang}, {Zhang},
  {Zhang}, {Zhang}, {Zhang}, {Zhang}, {Zhang}, {Zhang}, {Zhang}, {Zhang},
  {Zhang}, {Zhang}, {Zhang}, {Zhang}, {Zhang}, {Zhao}, {Zhao}, {Zhao}, {Zhao},
  {Zhao}, {Zheng}, {Zheng}, {Zhou}, {Zhou}, {Zhou}, {Zhou}, {Zhou}, {Zhou},
  {Zhou}, {Zhu}, {Zhu}, {Zhu}, {Zhu}, {Zou}, \& {Zuo}}]{2024arXiv240507691C}
---. 2024{\natexlab{b}}, arXiv e-prints, arXiv:2405.07691,
  \dodoi{10.48550/arXiv.2405.07691}

\bibitem[{{Cardullo} {et~al.}(2009){Cardullo}, {Corsini}, {Beifiori}, {Buson},
  {Dalla Bont{\`a}}, {Morelli}, {Pizzella}, \& {Bertola}}]{2009A&A...508..641C}
{Cardullo}, A., {Corsini}, E.~M., {Beifiori}, A., {et~al.} 2009, \aap, 508,
  641, \dodoi{10.1051/0004-6361/200913046}

\bibitem[{{Condon} {et~al.}(1998){Condon}, {Yin}, {Thuan}, \&
  {Boller}}]{1998AJ....116.2682C}
{Condon}, J.~J., {Yin}, Q.~F., {Thuan}, T.~X., \& {Boller}, T. 1998, \aj, 116,
  2682, \dodoi{10.1086/300624}

\bibitem[{{D'Abrusco} {et~al.}(2014){D'Abrusco}, {Fabbiano}, \&
  {Brassington}}]{2014ApJ...783...19D}
{D'Abrusco}, R., {Fabbiano}, G., \& {Brassington}, N.~J. 2014, \apj, 783, 19,
  \dodoi{10.1088/0004-637X/783/1/19}

\bibitem[{{D'Ammando}(2019)}]{2019Galax...7...87D}
{D'Ammando}, F. 2019, Galaxies, 7, 87, \dodoi{10.3390/galaxies7040087}

\bibitem[{{Di Gesu} {et~al.}(2022){Di Gesu}, {Donnarumma}, {Tavecchio},
  {Agudo}, {Barnounin}, {Cibrario}, {Di Lalla}, {Di Marco}, {Escudero},
  {Errando}, {Jorstad}, {Kim}, {Kouch}, {Liodakis}, {Lindfors}, {Madejski},
  {Marshall}, {Marscher}, {Middei}, {Muleri}, {Myserlis}, {Negro}, {Omodei},
  {Pacciani}, {Paggi}, {Perri}, {Puccetti}, {Antonelli}, {Bachetti}, {Baldini},
  {Baumgartner}, {Bellazzini}, {Bianchi}, {Bongiorno}, {Bonino}, {Brez},
  {Bucciantini}, {Capitanio}, {Castellano}, {Cavazzuti}, {Ciprini}, {Costa},
  {De Rosa}, {Del Monte}, {Doroshenko}, {Dov{\v{c}}iak}, {Ehlert}, {Enoto},
  {Evangelista}, {Fabiani}, {Ferrazzoli}, {Garcia}, {Gunji}, {Hayashida},
  {Heyl}, {Iwakiri}, {Karas}, {Kitaguchi}, {Kolodziejczak}, {Krawczynski}, {La
  Monaca}, {Latronico}, {Maldera}, {Manfreda}, {Marin}, {Marinucci}, {Massaro},
  {Matt}, {Mitsuishi}, {Mizuno}, {Ng}, {O'Dell}, {Oppedisano}, {Papitto},
  {Pavlov}, {Peirson}, {Pesce-Rollins}, {Petrucci}, {Pilia}, {Possenti},
  {Poutanen}, {Ramsey}, {Rankin}, {Ratheesh}, {Romani}, {Sgr{\`o}}, {Slane},
  {Soffitta}, {Spandre}, {Tamagawa}, {Taverna}, {Tawara}, {Tennant}, {Thomas},
  {Tombesi}, {Trois}, {Tsygankov}, {Turolla}, {Vink}, {Weisskopf}, {Wu}, {Xie},
  \& {Zane}}]{2022ApJ...938L...7D}
{Di Gesu}, L., {Donnarumma}, I., {Tavecchio}, F., {et~al.} 2022, \apjl, 938,
  L7, \dodoi{10.3847/2041-8213/ac913a}

\bibitem[{{Dutta} \& {Gupta}(2024)}]{2024arXiv240515657D}
{Dutta}, S., \& {Gupta}, N. 2024, arXiv e-prints, arXiv:2405.15657,
  \dodoi{10.48550/arXiv.2405.15657}

\bibitem[{{Fabbiano} {et~al.}(2010){Fabbiano}, {Brassington}, {Lentati},
  {Angelini}, {Davies}, {Gallagher}, {Kalogera}, {Kim}, {King}, {Kundu},
  {Pellegrini}, {Richings}, {Trinchieri}, {Zezas}, \&
  {Zepf}}]{2010ApJ...725.1824F}
{Fabbiano}, G., {Brassington}, N.~J., {Lentati}, L., {et~al.} 2010, \apj, 725,
  1824, \dodoi{10.1088/0004-637X/725/2/1824}

\bibitem[{{Finke} {et~al.}(2022){Finke}, {Ajello}, {Dom{\'\i}nguez}, {Desai},
  {Hartmann}, {Paliya}, \& {Saldana-Lopez}}]{2022ApJ...941...33F}
{Finke}, J.~D., {Ajello}, M., {Dom{\'\i}nguez}, A., {et~al.} 2022, \apj, 941,
  33, \dodoi{10.3847/1538-4357/ac9843}

\bibitem[{{Franceschini} \& {Rodighiero}(2017)}]{2017A&A...603A..34F}
{Franceschini}, A., \& {Rodighiero}, G. 2017, \aap, 603, A34,
  \dodoi{10.1051/0004-6361/201629684}

\bibitem[{{Fu} {et~al.}(2022){Fu}, {Zhang}, {Zhang}, {Liang}, {Yao}, \&
  {Liang}}]{2022RAA....22c5005F}
{Fu}, W.-J., {Zhang}, H.-M., {Zhang}, J., {et~al.} 2022, Research in Astronomy
  and Astrophysics, 22, 035005, \dodoi{10.1088/1674-4527/ac4410}

\bibitem[{{Funk} {et~al.}(2013){Funk}, {Hinton}, \& {CTA
  Consortium}}]{2013APh....43..348F}
{Funk}, S., {Hinton}, J.~A., \& {CTA Consortium}. 2013, Astroparticle Physics,
  43, 348, \dodoi{10.1016/j.astropartphys.2012.05.018}

\bibitem[{{Gan} {et~al.}(2024){Gan}, {Zhang}, {Yang}, {Gu}, \&
  {Zhang}}]{2024RAA....24b5018G}
{Gan}, Y.-Y., {Zhang}, H.-M., {Yang}, X., {Gu}, Y., \& {Zhang}, J. 2024,
  Research in Astronomy and Astrophysics, 24, 025018,
  \dodoi{10.1088/1674-4527/ad1c78}

\bibitem[{{Gan} {et~al.}(2021){Gan}, {Zhang}, {Zhang}, {Yang}, {Yi}, {Liang},
  \& {Liang}}]{2021RAA....21..201G}
{Gan}, Y.-Y., {Zhang}, H.-M., {Zhang}, J., {et~al.} 2021, Research in Astronomy
  and Astrophysics, 21, 201, \dodoi{10.1088/1674-4527/21/8/201}

\bibitem[{{Gan} {et~al.}(2022){Gan}, {Zhang}, {Yao}, {Zhang}, {Liang}, \&
  {Liang}}]{2022ApJ...939...78G}
{Gan}, Y.-Y., {Zhang}, J., {Yao}, S., {et~al.} 2022, \apj, 939, 78,
  \dodoi{10.3847/1538-4357/ac9589}

\bibitem[{{Giroletti} {et~al.}(2005){Giroletti}, {Taylor}, \&
  {Giovannini}}]{2005ApJ...622..178G}
{Giroletti}, M., {Taylor}, G.~B., \& {Giovannini}, G. 2005, \apj, 622, 178,
  \dodoi{10.1086/427898}

\bibitem[{{Gonz{\'a}lez-Mart{\'\i}n} {et~al.}(2009){Gonz{\'a}lez-Mart{\'\i}n},
  {Masegosa}, {M{\'a}rquez}, {Guainazzi}, \&
  {Jim{\'e}nez-Bail{\'o}n}}]{2009A&A...506.1107G}
{Gonz{\'a}lez-Mart{\'\i}n}, O., {Masegosa}, J., {M{\'a}rquez}, I., {Guainazzi},
  M., \& {Jim{\'e}nez-Bail{\'o}n}, E. 2009, \aap, 506, 1107,
  \dodoi{10.1051/0004-6361/200912288}

\bibitem[{{Grandi} {et~al.}(2016){Grandi}, {Capetti}, \&
  {Baldi}}]{2016MNRAS.457....2G}
{Grandi}, P., {Capetti}, A., \& {Baldi}, R.~D. 2016, \mnras, 457, 2,
  \dodoi{10.1093/mnras/stv2846}

\bibitem[{{Greenhill} {et~al.}(2003){Greenhill}, {Booth}, {Ellingsen},
  {Herrnstein}, {Jauncey}, {McCulloch}, {Moran}, {Norris}, {Reynolds}, \&
  {Tzioumis}}]{2003ApJ...590..162G}
{Greenhill}, L.~J., {Booth}, R.~S., {Ellingsen}, S.~P., {et~al.} 2003, \apj,
  590, 162, \dodoi{10.1086/374862}

\bibitem[{{Gu} {et~al.}(2022){Gu}, {Zhang}, {Gan}, {Zhang}, {Sun}, \&
  {Liang}}]{2022ApJ...927..221G}
{Gu}, Y., {Zhang}, H.-M., {Gan}, Y.-Y., {et~al.} 2022, \apj, 927, 221,
  \dodoi{10.3847/1538-4357/ac540e}

\bibitem[{{H.~E.~S.~S. Collaboration} {et~al.}(2015){H.~E.~S.~S.
  Collaboration}, {Abramowski}, {Aharonian}, {Ait Benkhali}, {Akhperjanian},
  {Ang{\"u}ner}, {Backes}, {Balzer}, {Becherini}, {Becker Tjus}, {Berge},
  {Bernhard}, {Bernl{\"o}hr}, {Birsin}, {Blackwell}, {B{\"o}ttcher}, {Boisson},
  {Bolmont}, {Bordas}, {Bregeon}, {Brun}, {Brun}, {Bryan}, {Bulik}, {Carr},
  {Casanova}, {Chakraborty}, {Chalme-Calvet}, {Chaves}, {Chen}, {Chr{\'e}tien},
  {Colafrancesco}, {Cologna}, {Conrad}, {Couturier}, {Cui}, {Davids},
  {Degrange}, {Deil}, {deWilt}, {Djannati-Ata{\"\i}}, {Domainko}, {Donath},
  {O'C. Drury}, {Dubus}, {Dutson}, {Dyks}, {Dyrda}, {Edwards}, {Egberts},
  {Eger}, {Ernenwein}, {Espigat}, {Farnier}, {Fegan}, {Feinstein}, {Fernandes},
  {Fernandez}, {Fiasson}, {Fontaine}, {F{\"o}rster}, {F{\"u}{\ss}ling},
  {Gabici}, {Gajdus}, {Gallant}, {Garrigoux}, {Giavitto}, {Giebels},
  {Glicenstein}, {Gottschall}, {Goyal}, {Grondin}, {Grudzi{\'n}ska}, {Hadasch},
  {H{\"a}ffner}, {Hahn}, {Hawkes}, {Heinzelmann}, {Henri}, {Hermann}, {Hervet},
  {Hillert}, {Hinton}, {Hofmann}, {Hofverberg}, {Hoischen}, {Holler}, {Horns},
  {Ivascenko}, {Jacholkowska}, {Jahn}, {Jamrozy}, {Janiak}, {Jankowsky},
  {Jung-Richardt}, {Kastendieck}, {Katarzy{\'n}ski}, {Katz}, {Kerszberg},
  {Kh{\'e}lifi}, {Kieffer}, {Klepser}, {Klochkov}, {Klu{\'z}niak}, {Kolitzus},
  {Komin}, {Kosack}, {Krakau}, {Krayzel}, {Kr{\"u}ger}, {Laffon}, {Lamanna},
  {Lau}, {Lefaucheur}, {Lefranc}, {Lemi{\`e}re}, {Lemoine-Goumard}, {Lenain},
  {Lohse}, {Lopatin}, {Lu}, {Lui}, {Marandon}, {Marcowith}, {Mariaud}, {Marx},
  {Maurin}, {Maxted}, {Mayer}, {Meintjes}, {Menzler}, {Meyer}, {Mitchell},
  {Moderski}, {Mohamed}, {Mor{\r{a}}}, {Moulin}, {Murach}, {de Naurois},
  {Niemiec}, {Oakes}, {Odaka}, {{\"O}ttl}, {Ohm}, {de O{\~n}a Wilhelmi},
  {Opitz}, {Ostrowski}, {Oya}, {Panter}, {Parsons}, {Arribas}, {Pekeur},
  {Pelletier}, {Petrucci}, {Peyaud}, {Pita}, {Poon}, {Prokoph},
  {P{\"u}hlhofer}, {Punch}, {Quirrenbach}, {Raab}, {Reichardt}, {Reimer},
  {Reimer}, {Renaud}, {de los Reyes}, {Rieger}, {Romoli}, {Rosier-Lees},
  {Rowell}, {Rudak}, {Rulten}, {Sahakian}, {Salek}, {Sanchez}, {Santangelo},
  {Sasaki}, {Schlickeiser}, {Sch{\"u}ssler}, {Schulz}, {Schwanke}, {Schwemmer},
  {Seyffert}, {Simoni}, {Sol}, {Spanier}, {Spengler}, {Spies}, {Stawarz},
  {Steenkamp}, {Stegmann}, {Stinzing}, {Stycz}, {Sushch}, {Tavernet},
  {Tavernier}, {Taylor}, {Terrier}, {Tluczykont}, {Trichard}, {Valerius}, {van
  der Walt}, {van Eldik}, {van Soelen}, {Vasileiadis}, {Veh}, {Venter},
  {Viana}, {Vincent}, {Vink}, {Voisin}, {V{\"o}lk}, {Vuillaume}, {Wagner},
  {Wagner}, {Wagner}, {Weidinger}, {Weitzel}, {White}, {Wierzcholska},
  {Willmann}, {W{\"o}rnlein}, {Wouters}, {Yang}, {Zabalza}, {Zaborov},
  {Zacharias}, {Zdziarski}, {Zech}, {Zefi}, \&
  {{\.Z}ywucka}}]{2015A&A...577A.131H}
{H.~E.~S.~S. Collaboration}, {Abramowski}, A., {Aharonian}, F., {et~al.} 2015,
  \aap, 577, A131, \dodoi{10.1051/0004-6361/201525699}

\bibitem[{{Harvey} {et~al.}(2022){Harvey}, {Rulten}, \&
  {Chadwick}}]{2022MNRAS.512.1141H}
{Harvey}, M., {Rulten}, C.~B., \& {Chadwick}, P.~M. 2022, \mnras, 512, 1141,
  \dodoi{10.1093/mnras/stac375}

\bibitem[{{Heckman}(1980)}]{1980A&A....87..152H}
{Heckman}, T.~M. 1980, \aap, 87, 152

\bibitem[{{Hu} {et~al.}(2024){Hu}, {Yu}, {Zhang}, {Wang}, {Patra}, {Brink},
  {Zheng}, {Wang}, {Kong}, {Chen}, {Zhou}, {Cao}, {Lu}, {Zhou}, {Wei}, {Huang},
  {Li}, {Lou}, {Mao}, {Liang}, \& {Filippenko}}]{2024ApJ...970L..22H}
{Hu}, X.-K., {Yu}, Y.-W., {Zhang}, J., {et~al.} 2024, \apjl, 970, L22,
  \dodoi{10.3847/2041-8213/ad5e68}

\bibitem[{{Kalberla} {et~al.}(2005){Kalberla}, {Burton}, {Hartmann}, {Arnal},
  {Bajaja}, {Morras}, \& {P{\"o}ppel}}]{2005A&A...440..775K}
{Kalberla}, P.~M.~W., {Burton}, W.~B., {Hartmann}, D., {et~al.} 2005, \aap,
  440, 775, \dodoi{10.1051/0004-6361:20041864}

\bibitem[{{Kantzas} {et~al.}(2022){Kantzas}, {Markoff}, {Lucchini},
  {Ceccobello}, {Grinberg}, {Connors}, \& {Uttley}}]{2022MNRAS.510.5187K}
{Kantzas}, D., {Markoff}, S., {Lucchini}, M., {et~al.} 2022, \mnras, 510, 5187,
  \dodoi{10.1093/mnras/stac004}

\bibitem[{{Kiehlmann} {et~al.}(2024){Kiehlmann}, {Lister}, {Readhead},
  {Liodakis}, {O'Neill}, {Pearson}, {Sheldahl}, {Siemiginowska}, {Tassis},
  {Taylor}, \& {Wilkinson}}]{2024ApJ...961..240K}
{Kiehlmann}, S., {Lister}, M.~L., {Readhead}, A.~C.~S., {et~al.} 2024, \apj,
  961, 240, \dodoi{10.3847/1538-4357/ad0c56}

\bibitem[{{Kneiske} \& {Raue}(2010)}]{2010cosp...38.2365K}
{Kneiske}, T., \& {Raue}, M. 2010, in 38th COSPAR Scientific Assembly, Vol.~38,
  3

\bibitem[{{Lanzetta} {et~al.}(1993){Lanzetta}, {Turnshek}, \&
  {Sandoval}}]{1993ApJS...84..109L}
{Lanzetta}, K.~M., {Turnshek}, D.~A., \& {Sandoval}, J. 1993, \apjs, 84, 109,
  \dodoi{10.1086/191749}

\bibitem[{{Li} {et~al.}(2014){Li}, {Kong}, {Takata}, {Cheng}, {Tam}, {Hui}, \&
  {Jin}}]{2014ApJ...797..111L}
{Li}, K.~L., {Kong}, A.~K.~H., {Takata}, J., {et~al.} 2014, \apj, 797, 111,
  \dodoi{10.1088/0004-637X/797/2/111}

\bibitem[{{Lister} {et~al.}(2020){Lister}, {Homan}, {Kovalev}, {Mandal},
  {Pushkarev}, \& {Siemiginowska}}]{2020ApJ...899..141L}
{Lister}, M.~L., {Homan}, D.~C., {Kovalev}, Y.~Y., {et~al.} 2020, \apj, 899,
  141, \dodoi{10.3847/1538-4357/aba18d}

\bibitem[{{Ma} {et~al.}(2022){Ma}, {Bi}, {Cao}, {Chen}, {Chen}, {Cheng},
  {Gong}, {Gu}, {He}, {Hou}, {Huang}, {Huang}, {Liu}, {Shchegolev}, {Sheng},
  {Stenkin}, {Wu}, {Wu}, {Wu}, {Xiao}, {Yao}, {Zhang}, {Zhang}, \&
  {Zuo}}]{2022ChPhC..46c0001M}
{Ma}, X.-H., {Bi}, Y.-J., {Cao}, Z., {et~al.} 2022, Chinese Physics C, 46,
  030001, \dodoi{10.1088/1674-1137/ac3fa6}

\bibitem[{{Mattox} {et~al.}(1996){Mattox}, {Bertsch}, {Chiang}, {Dingus},
  {Digel}, {Esposito}, {Fierro}, {Hartman}, {Hunter}, {Kanbach}, {Kniffen},
  {Lin}, {Macomb}, {Mayer-Hasselwander}, {Michelson}, {von Montigny},
  {Mukherjee}, {Nolan}, {Ramanamurthy}, {Schneid}, {Sreekumar}, {Thompson}, \&
  {Willis}}]{1996ApJ...461..396M}
{Mattox}, J.~R., {Bertsch}, D.~L., {Chiang}, J., {et~al.} 1996, \apj, 461, 396,
  \dodoi{10.1086/177068}

\bibitem[{{Migliori} {et~al.}(2016){Migliori}, {Siemiginowska}, {Sobolewska},
  {Loh}, {Corbel}, {Ostorero}, \& {Stawarz}}]{2016agnt.confE..60M}
{Migliori}, G., {Siemiginowska}, A., {Sobolewska}, M., {et~al.} 2016, in Active
  Galactic Nuclei 12: A Multi-Messenger Perspective (AGN12), 60,
  \dodoi{10.5281/zenodo.163821}

\bibitem[{{O'Dea} \& {Saikia}(2021)}]{2021A&ARv..29....3O}
{O'Dea}, C.~P., \& {Saikia}, D.~J. 2021, \aapr, 29, 3,
  \dodoi{10.1007/s00159-021-00131-w}

\bibitem[{{Paliya}(2019)}]{2019JApA...40...39P}
{Paliya}, V.~S. 2019, Journal of Astrophysics and Astronomy, 40, 39,
  \dodoi{10.1007/s12036-019-9604-3}

\bibitem[{{Principe} {et~al.}(2020){Principe}, {Migliori}, {Johnson},
  {D'Ammando}, {Giroletti}, {Orienti}, {Stanghellini}, {Taylor}, {Torresi}, \&
  {Cheung}}]{2020A&A...635A.185P}
{Principe}, G., {Migliori}, G., {Johnson}, T.~J., {et~al.} 2020, \aap, 635,
  A185, \dodoi{10.1051/0004-6361/201937049}

\bibitem[{{Scarpa} \& {Falomo}(1997)}]{1997A&A...325..109S}
{Scarpa}, R., \& {Falomo}, R. 1997, \aap, 325, 109

\bibitem[{{Sikora} {et~al.}(2007){Sikora}, {Stawarz}, \&
  {Lasota}}]{2007ApJ...658..815S}
{Sikora}, M., {Stawarz}, {\L}., \& {Lasota}, J.-P. 2007, \apj, 658, 815,
  \dodoi{10.1086/511972}

\bibitem[{{Sun} {et~al.}(2015){Sun}, {Zhang}, {Lin}, {Xue}, {Liang}, \&
  {Zhang}}]{2015ApJ...798...43S}
{Sun}, X.-N., {Zhang}, J., {Lin}, D.-B., {et~al.} 2015, \apj, 798, 43,
  \dodoi{10.1088/0004-637X/798/1/43}

\bibitem[{{Tavecchio} {et~al.}(2010){Tavecchio}, {Ghisellini}, {Ghirlanda},
  {Foschini}, \& {Maraschi}}]{2010MNRAS.401.1570T}
{Tavecchio}, F., {Ghisellini}, G., {Ghirlanda}, G., {Foschini}, L., \&
  {Maraschi}, L. 2010, \mnras, 401, 1570,
  \dodoi{10.1111/j.1365-2966.2009.15784.x}

\bibitem[{{Tramacere} {et~al.}(2009){Tramacere}, {Giommi}, {Perri},
  {Verrecchia}, \& {Tosti}}]{2009A&A...501..879T}
{Tramacere}, A., {Giommi}, P., {Perri}, M., {Verrecchia}, F., \& {Tosti}, G.
  2009, \aap, 501, 879, \dodoi{10.1051/0004-6361/200810865}

\bibitem[{{Tramacere} {et~al.}(2011){Tramacere}, {Massaro}, \&
  {Taylor}}]{2011ApJ...739...66T}
{Tramacere}, A., {Massaro}, E., \& {Taylor}, A.~M. 2011, \apj, 739, 66,
  \dodoi{10.1088/0004-637X/739/2/66}

\bibitem[{{Urry} \& {Padovani}(1995)}]{1995PASP..107..803U}
{Urry}, C.~M., \& {Padovani}, P. 1995, \pasp, 107, 803, \dodoi{10.1086/133630}

\bibitem[{{V{\'e}ron-Cetty} \& {V{\'e}ron}(2003)}]{2003A&A...412..399V}
{V{\'e}ron-Cetty}, M.~P., \& {V{\'e}ron}, P. 2003, \aap, 412, 399,
  \dodoi{10.1051/0004-6361:20034225}

\bibitem[{{Wang} {et~al.}(2022){Wang}, {Bi}, {Cao}, {Vallania}, {Wu}, {Yan}, \&
  {Yuan}}]{2022ChPhC..46c0003W}
{Wang}, X.-Y., {Bi}, X.-J., {Cao}, Z., {et~al.} 2022, Chinese Physics C, 46,
  030003, \dodoi{10.1088/1674-1137/ac3fa9}

\bibitem[{{Wang} {et~al.}(2024){Wang}, {Xue}, {Xiong}, {Wang}, {Sun}, {Peng},
  \& {Mao}}]{2024ApJS..271...10W}
{Wang}, Z.-R., {Xue}, R., {Xiong}, D., {et~al.} 2024, \apjs, 271, 10,
  \dodoi{10.3847/1538-4365/ad168c}

\bibitem[{{Wood} {et~al.}(2017){Wood}, {Caputo}, {Charles}, {Di Mauro},
  {Magill}, {Perkins}, \& {Fermi-LAT Collaboration}}]{Wood2017}
{Wood}, M., {Caputo}, R., {Charles}, E., {et~al.} 2017, in International Cosmic
  Ray Conference, Vol. 301, 35th International Cosmic Ray Conference
  (ICRC2017), 824, \dodoi{10.22323/1.301.0824}

\bibitem[{{Xing} \& {Wang}(2015)}]{2015ApJ...808...17X}
{Xing}, Y., \& {Wang}, Z. 2015, \apj, 808, 17,
  \dodoi{10.1088/0004-637X/808/1/17}

\bibitem[{{Xing} {et~al.}(2018){Xing}, {Wang}, \&
  {Takata}}]{2018RAA....18..127X}
{Xing}, Y., {Wang}, Z.-X., \& {Takata}, J. 2018, Research in Astronomy and
  Astrophysics, 18, 127, \dodoi{10.1088/1674-4527/18/10/127}

\bibitem[{{Yan} {et~al.}(2014){Yan}, {Zeng}, \& {Zhang}}]{2014MNRAS.439.2933Y}
{Yan}, D., {Zeng}, H., \& {Zhang}, L. 2014, \mnras, 439, 2933,
  \dodoi{10.1093/mnras/stu146}

\bibitem[{{Younes} {et~al.}(2010){Younes}, {Porquet}, {Sabra}, {Grosso},
  {Reeves}, \& {Allen}}]{2010A&A...517A..33Y}
{Younes}, G., {Porquet}, D., {Sabra}, B., {et~al.} 2010, \aap, 517, A33,
  \dodoi{10.1051/0004-6361/201014371}

\bibitem[{{Yu} {et~al.}(2024){Yu}, {Zhang}, {Gan}, {Hu}, {Wu}, \&
  {Zhang}}]{2024ApJ...965..163Y}
{Yu}, Y.-W., {Zhang}, H.-M., {Gan}, Y.-Y., {et~al.} 2024, \apj, 965, 163,
  \dodoi{10.3847/1538-4357/ad2e07}

\bibitem[{{Zhang} {et~al.}(2012){Zhang}, {Liang}, {Zhang}, \&
  {Bai}}]{2012ApJ...752..157Z}
{Zhang}, J., {Liang}, E.-W., {Zhang}, S.-N., \& {Bai}, J.~M. 2012, \apj, 752,
  157, \dodoi{10.1088/0004-637X/752/2/157}

\bibitem[{{Zhang} {et~al.}(2020){Zhang}, {Zhang}, {Gan}, {Yi}, {Wang}, \&
  {Liang}}]{2020ApJ...899....2Z}
{Zhang}, J., {Zhang}, H.-M., {Gan}, Y.-Y., {et~al.} 2020, \apj, 899, 2,
  \dodoi{10.3847/1538-4357/aba2cd}

\bibitem[{{Zhang} {et~al.}(2013){Zhang}, {Zhang}, \&
  {Liang}}]{2013ApJ...767....8Z}
{Zhang}, J., {Zhang}, S.-N., \& {Liang}, E.-W. 2013, \apj, 767, 8,
  \dodoi{10.1088/0004-637X/767/1/8}

\end{thebibliography}
\bibliographystyle{aasjournal}
\clearpage

\begin{table*}[htbp]
    \begin{center}
    \caption{Observations of {\it Chandra} and {\it Swift}-XRT and Data Analysis Results for NGC 4278}
    \label{tab:X-ray}
    \begin{tabular}{ccccccccc}
    \hline\hline
     Obs-date & Obs-ID &Exposure & $N^{\rm H}_{\rm int}$ &$\Gamma_{\rm X}$ &PL norm & kT & Corr.flux & Refs\tablenotemark{*}
     \\
    ~   &  ~  &  $[\rm ks]$  &$[10^{20}\rm cm^{-2}]$ & ~ & $[\rm 10^{-4}\ Ph\ keV^{-1}\ cm^{-2}\ s^{-1}]$ & $[\rm keV]$ & $[\rm 10^{-13}\ erg\ cm^{-2}\ s^{-1}]$&  \\
    \hline
    2000-04-20& 398  &  1.4  &         &                        &                        &                &                         &a \\
    2005-02-03& 4741 & 37.5  & $<6.78$ & $2.13^{+0.15}_{-0.13}$ & $4.28^{+0.43}_{-0.36}$ & $0.62\pm 0.04$ &$18.5^{+1.6}_{-0.3}$     &a \\
    2006-03-16& 7077 & 110.3 &         & $2.26^{+0.13}_{-0.10}$ & $1.82^{+0.17}_{-0.14}$ &                &$8.21^{+0.79}_{-0.21}$   &a \\
    2006-07-25& 7078 &  51.4 &         & $2.34^{+0.13}_{-0.12}$ & $4.21^{+0.40}_{-0.31}$ &                &$16.5^{+2.0}_{-0.4}$     &a \\
    2006-10-24& 7079 & 105.1 &         & $2.38^{+0.12}_{-0.10}$ & $3.76^{+0.31}_{-0.22}$ &                &$15.05^{+1.34}_{-0.35}$  &a \\
    2007-04-20& 7080 & 55.8  &         & $2.02^{+0.18}_{-0.18}$ & $1.10^{+0.16}_{-0.14}$ &                &$6.0^{+0.8}_{-0.3}$      &a \\
    2007-02-20& 7081 & 110.7 &         & $2.12^{+0.13}_{-0.12}$ & $1.25^{+0.04}_{-0.11}$ &                &$6.4^{+0.7}_{-0.2}$      &a \\  
    \hline   
    2010-03-15& 11269 & 81.9 & $<2.43$ & $2.14^{+0.13}_{-0.11}$ & $0.35^{+0.04}_{-0.03}$ & $0.62^{+0.04}_{-0.05}$ & $1.86\pm 0.06$  &b \\   
    2010-03-20& 12124 & 25.8 & $<11.62$& $2.29^{+0.41}_{-0.22}$ & $0.33^{+0.12}_{-0.08}$ & $0.57^{+0.10}_{-0.16}$ & $1.66\pm 0.11$  &b \\
    \hline   
    2021      &       &      &$2.43$   & $1.65^{+0.37}_{-0.36}$ & $5.39^{+1.53}_{-1.29}$ &                &$31.86^{+6.09}_{-6.51}$  &c \\
    \hline
    \hline
    \end{tabular}
    \end{center}
    \tablenotetext{*}{`a' denotes the {\it Chandra} observation results in \cite{2010A&A...517A..33Y}, while `b' and `c' respectively represent the analysis results of {\it Chandra} and {\it Swift}-XRT observations conducted in this work.}
\end{table*}

\begin{table*}
    \begin{center}
    \caption{SED Fitting Parameters and Derived Jet Powers of NGC 4278}
    \label{tab:paras}
    \begin{tabular}{ccccccccccccccccc}
    \hline\hline
   $N_0$    & $p_1^{\star}$ &  $p_2^{\star}$ & $\gamma_{\rm min}^{\star}$ & $\gamma_{\rm b}$ & $\gamma_{\rm max}^{\star}$ & $B$ & $\delta^{\star}$ & $R^{\star}$ & $P_{\rm e}$ & $P_{\rm B}$ & $P_{\rm r}$ & $P_{\rm jet}$ & Refs\tablenotemark{*}\\  
    $[\rm 1/cm^3]$ & ~  & ~ & ~ & ~ & ~  & [\rm mG] & ~ & $[\rm cm]$ & $[\rm erg\ s^{-1}]$ & $[\rm erg\ s^{-1}]$ & $[\rm erg\ s^{-1}]$ & $[\rm erg\ s^{-1}]$ &  \\
    \hline
    4.5E4 & 2.27  & 3.8 & 1    & 4E6  &  3E7   & 7  & 2.7   & 1E16   &1.84E43  &2.68E38  & 1.69E41 & 1.86E43 & a \\  
    2.1E1 & 1.70   &  3.0  & 1    & 3E5  &  3E7   &1 & 10    & 1E16   & 6.05E42 & 1.08E38 & 1.10E40 & 6.06E42 & a \\  \hline
        ~ & $1$   & $3.0$ &1E2   &9E5   &  1E7   &40   & 9.7   & 8.5E13 & ~       & ~       & ~       & ~       & b1          \\  
        ~ & $1$   &$2.3$& 1E2  &3E6   &  5E7   &200   & 1.8   & 1.5E14 & ~       & ~       & ~       & ~       &b2         \\      
        ~ & ~     &  ~  &6.2E3 & ~    & 1.9E8  &8   & $2.7$ & 6E15   & ~       & ~       & ~       & ~       &    c                  \\    
    \hline
    \hline
    \end{tabular}
    \end{center}
    \tablenotetext{\star}{The parameters remain fixed during SED modeling.}
     \tablenotetext{*}{`a' denotes the parameter values in this work,`b1' and `b2' represent the fitting parameters in \cite{2024ApJS..271...10W} for viewing angles of $\theta=1.8\degr$ and $\theta=30\degr$, respectively, while `c' indicates the results in \cite{2024arXiv240515657D} by considering a Log-Parabola electron spectrum.}

\end{table*}

\clearpage

\end{document}